\documentclass[12pt,preprint]{aastex}
\usepackage{multirow,amsthm,amsmath,amssymb,mathrsfs,ulem}
\usepackage{graphicx,subfigure,verbatim,color,soul,lineno}
\graphicspath{{./}}

\begin{document}

\title{Galactic diffuse gamma-ray emission from GeV to PeV energies 
in light of up-to-date cosmic ray measurements}

\author{Rui Zhang$^{1,2}$, Xiaoyuan Huang$^{1,2}$, Zhi-Hui Xu$^{3}$, 
Shiping Zhao$^{1}$, Qiang Yuan$^{1,2}$}

\affil{$^1$Key Laboratory of Dark Matter and Space Astronomy, Purple Mountain Observatory, Chinese Academy of Sciences, Nanjing 210023, China; xyhuang@pmo.ac.cn, yuanq@pmo.ac.cn \\
$^2$School of Astronomy and Space Science, University of Science and Technology of China, Hefei 230026, China\\
$^3$Institute of Modern Physics, Chinese Academy of Sciences, Lanzhou 730000, China
}

\begin{abstract}
The diffuse $\gamma$-ray emission between 10 and 1000 TeV from the Galactic plane 
was recently measured by the Large High Altitude Air Shower Observatory (LHAASO). 
 These observations will help tremendously in constraining the propagation and 
interaction of cosmic rays in the Milky Way. Additionally, new measurements of CR 
spectra reach a very high precision up to 100 TeV energies, revealing multiple spectral
structures of various species. In this work, we confront the model prediction of the 
diffuse $\gamma$-ray emission, based on up-to-date measurements of the local cosmic ray 
spectra and simplified propagation setup, with the measurements of diffuse $\gamma$-rays. 
To better constrain the low-energy part of the model, we analyze the 14.6 years 
of Fermi-LAT data to extract the Galactic diffuse emission between 1 and 500 GeV 
from the same sky regions of LHAASO, after subtracting the contribution from known 
sources and the isotropic diffuse $\gamma$-ray background. The joint Fermi-LAT and
LHAASO spectra thus cover a very wide energy range from 1 GeV to 1 PeV with small
gaps from 0.5 to 10 TeV.
Compared with the prediction, we find that clear excesses between several GeV and 
$\sim 60$ TeV of the diffuse emission exist. Possible reasons to explain the excesses
may include unresolved sources or more complicated propagation models. We illustrate
that an exponential-cutoff-power-law component with an index of $-2.40$ and cutoff 
energy of $\sim30$ TeV is able to account for such excesses. 
\end{abstract}

\section{Introduction}

The origin and propagation of cosmic rays (CRs) are one of the most important 
questions in astroparticle physics. Except for the ultra-high energy end ($E\gtrsim1$~EeV), 
most CR particles lose their directions due to the deflection in the Galactic 
magnetic field, resulting in difficulties in understanding their origin.
Because measurements of CRs only exist for the location of the Sun, 
the overall distribution of CRs in the Milky Way can only be inferred indirectly.
The diffuse $\gamma$-ray emission produced by interactions between CR
particles and the diffuse medium and radiation fields, on the other hand, 
carries important information of the distribution of CRs far away from 
the solar location and thus is crucial to constraining the origin and 
propagation of CRs \citep{2000ApJ...537..763S,2004ApJ...613..962S,
2007ARNPS..57..285S}.

In past years, both the (direct or indirect) measurements of CRs and diffuse 
$\gamma$ rays achieve big progresses.  
In direct detection experiments, carried out in space or on balloons, 
CR particles directly enter the detector. This allows to determine not only the 
energy and direction, but also to measure precisely the charge and mass numbers.
Hardening features around hundreds of GV rigidities 
\citep{2007BRASP..71..494P,2010ApJ...714L..89A,2011Sci...332...69A,
2015PhRvL.114q1103A,2015PhRvL.115u1101A,2019PhRvL.122r1102A,
2019SciA....5.3793A,2021PhRvL.126t1102A} 
and softening features around 10 TV rigidity 
\citep{2017ApJ...839....5Y,2018JETPL.108....5A,2019SciA....5.3793A,
2021PhRvL.126t1102A,2022PhRvL.129j1102A,2022ApJ...940..107C}
were revealed by many direct detection experiments.
The indirect measurements by ground-based experiments detect the secondary 
particles or photons produced when CR particles interact with the atmosphere. 
This method allows to detect CRs with even higher energies. However, a precise 
measurement of the particle type is not possible anymore, due to the relatively 
small differences of secondary particle contents of different primary nuclei 
with the same energy \citep{Engel:2011zzb}, and hence the spectra of 
individual particle species have relatively large uncertainties 
\citep{2005APh....24....1A,2013APh....47...54A,2019PhRvD.100h2002A}. 
Nevertheless, the understanding of the wide-band energy spectra below the so-called 
``knee'' of CRs\footnote{The “knee” is a spectral softening feature at several PeV, 
which may reflect the maximum acceleration energy of Galactic sources or the leakage 
of CRs to space out of the Galaxy \citep{2004APh....21..241H}.}, 
improves remarkably than before. As for the diffuse $\gamma$-ray 
emission, the space satellites measured the all-sky emission up to TeV energies
\citep{1997ApJ...481..205H,2012ApJ...750....3A}. 
It was shown that predictions, based on CR distribution and the diffuse medium 
and radiation fields, are consistent with the data measured by Fermi-LAT at high and 
intermediate latitudes but under-predict the data in the inner Galaxy for energies 
above a few GeV \citep{2012ApJ...750....3A}. At very-high-energy (VHE) to 
ultra-high-energy (UHE) bands, diffuse $\gamma$-ray emission were only detected from 
selected regions of the Galactic plane by ground-based experiments such as Milagro
\citep{2007ApJ...658L..33A,2008ApJ...688.1078A}, H.E.S.S. \citep{2014PhRvD..90l2007A}, 
ARGO-YBJ \citep{2015ApJ...806...20B}, and Tibet AS$\gamma$ \citep{2021PhRvL.126n1101A}. 
The HAWC experiment also reported a preliminary analysis of the diffuse emission from
the inner region of the Galactic plane \citep{HAWC:2021bvb}. Even if the flux 
points of the VHE diffuse emission are sparse and the uncertainties are large, 
these results shed new light on the understanding of the production and 
propagation of Galactic CRs \citep{2021ApJ...919...93F,2021ApJ...914L...7L,
2021PhRvD.104d3010K,2021Univ....7..141T,2022FrPhy..1744501Q,
2021arXiv210402838D,2022A&A...661A..72B,2022PhRvD.105b3002Z}.

Very recently, significantly improved measurements of the VHE-UHE diffuse 
emission from 10 TeV to 1000 TeV from both the inner and outer Galactic plane 
\citep{LHAASO-diffuse-2023} were obtained by the Large High Altitude Air 
Shower Observatory \citep[LHAASO;][]{2022ChPhC..46c0001M}.
Particularly, LHAASO measured the diffuse emission from the outer Galactic
plane for the first time, which can give more comprehensive constraints on 
the CR distribution in the Galaxy. Based on the improved measurements of the 
local CR spectra and diffuse emission, one can better constrain the 
CR injection, propagation, and interaction in the Milky Way 
\citep{2022arXiv221115607S,2023MNRAS.518.5036M}. 

In this work, we re-visit the diffuse $\gamma$-ray emission from the 
interactions between CRs and the interstellar medium (ISM) or the interstellar
radiation field (ISRF), according to the new measurements.
The injection and propagation parameters of CRs are tuned to fit the up-to-date 
measurements of primary and secondary CRs (including electrons and positrons) 
in a wide energy range. The measurements of low-energy spectra out of the 
canonical heliosphere by Voyager-1 are also used to constrain the CR spectra 
in the local interstellar space \citep{2016ApJ...831...18C}.
Diffuse $\gamma$-ray emission from CR interactions is then calculated and 
compared with the measurements from GeV to sub-PeV energies. To enable a 
consistent comparison between the model and data, we re-analyze the Fermi-LAT 
data in the same sky regions as the LHAASO measurements, and apply the same 
masks to reduce the contamination of resolved sources to the diffuse emission.

The rest of this paper is arranged as follows. In Sec. 2 we present the Fermi-LAT 
data analysis. In Sec. 3 we describe the model of diffuse $\gamma$-ray emission,
including the CR propagation model, the fitting of CR spectra, and the calculation 
of $\gamma$-ray production and absorption. In Sec. 4 we confront the model predictions 
of the diffuse emission with the wide-band data. We conclude this work in Sec. 5.

\section{Fermi-LAT data analysis}

To compare the model prediction of the Galactic diffuse emission with observations 
in a wide energy band, we analyze the Fermi-LAT data from 1 GeV to 500 GeV in 
selected sky regions same as those adopted in the LHAASO analysis, i.e., the inner 
Galaxy region with $-5^{\circ}<b<5^{\circ}$, $15^{\circ}<l<125^{\circ}$ 
and the outer Galaxy region with $-5^{\circ}<b<5^{\circ}$, 
$125^{\circ}<l<235^{\circ}$. The newest reconstructed P8R3 ULTRACLEANVETO Fermi-LAT 
data\footnote{https://fermi.gsfc.nasa.gov/ssc/data/access/}, which have a good 
CR background rejection performance, are used. 
We select in total 761 weeks of data recorded from August 4, 2008 to 
March 2, 2023. To suppress the contamination from $\gamma$-rays 
generated by CR interactions in the upper layers of the atmosphere, 
photons with zenith angles larger than 90$^\circ$ are removed. Moreover, 
we filter the data using the specification {\tt{(DATA$\_$QUAL$>$0) $\&\&$
(LAT$\_$CONFIG==1)}} to select good time intervals in which the satellite 
was working in the standard data-taking mode and the data quality is good. 
We bin the data into 15 logarithmically evenly distributed energy bins 
and take $0.1^\circ\times0.1^\circ$ pixel size for the spatial binning. 
%Since the DGE model {{\tt gll\_iem\_v07.fits}} provided by the Fermi Science 
%Support Center\footnote{http://fermi.gsfc.nasa.gov/ssc/data/access/lat/BackgroundModels.html}  
%has combined $\gamma$ rays emitted by different mechanisms into a single 
%file, it is not flexible enough for our usage. Thus we use DGE modelled by 
%the GALPROP software\footnote{http://galprop.stanford.edu/} with the Model 
%1, 2 and 3 introduced later. For the isotropic diffuse $\gamma$-ray 
%component, we do not use the default model {\tt iso\_P8R3\_SOURCE\_V2\_v1.txt} 
%accompanied with galactic diffuse model {\tt gll\_iem\_v07.fits}. 
Some of the sources detected by Fermi-LAT are not detected by KM2A, and we need 
to remove their contribution properly. The source model XML file is generated 
based on the 4FGL catalog \citep{2020ApJS..247...33A}. 
%taking sources within our region of interest considering the good performance 
%of the point spread function (PSF) in our selected energy band.
For the diffuse background emissions, we take the Galactic diffuse model 
{\tt gll\_iem\_v07.fits} and the isotropic background spectrum 
{\tt iso\_P8R3\_ULTRACLEANVETO\_V3\_v1.txt} as recommended by the Fermi-LAT
collaboration\footnote{https://fermi.gsfc.nasa.gov/ssc/data/access/lat/BackgroundModels.html}. 
We employ the binned likelihood analysis method to analyze the data with 
Fermitools {\tt version 2.2.0}, to re-fit the contribution from point sources 
and extended sources within our regions of interest. The instrument response 
function (IRF) adopted is {\tt P8R3\_ULTRACLEANVETO\_V3}. Then we take the 
best-fitting model and use {\tt gtmodel} to generate the model cube which 
includes contributions from point sources, extended sources, and the isotropic 
background. Through subtracting the model cube obtained above from the data, 
we get the residual photon counts which are expected to be mainly from the 
Galactic diffuse emission. To have consistent sky regions with the LHAASO 
measurements, we use the same mask method adopted in the LHAASO analysis
\citep{LHAASO-diffuse-2023}. We divide the residual photon counts by the 
energy interval of the energy bin, the mean exposure, and the solid angle 
of selected regions, to obtain the fluxes.

Due to the large photon counts observed by Fermi-LAT, the statistical errors 
are small. However, there are systematic errors which need to be considered.
Our target regions lie in the Galactic plane, therefore the contribution from the 
isotropic component is expected to be sub-dominant,  as indicated in 
\citet{2012ApJ...750....3A} for the Galactic disk region. We use different 
functional forms of the isotropic component given in \citet{2015ApJ...799...86A}, 
and find that this would affect the fluxes by about 2.5\%. We also checked the 
effect of different GDE models on the resulting fluxes, and it is negligible.
%However, for the GDE model, different models from \citet{2015ApJ...799...86A}  
%would only have a negligible effect on the resulting fluxes, since we obtain 
%them from residual photon counts. 
Since our target regions cover a large portion of the sky, the exposure at 
different energies in these areas may vary at the 10\% level. The systematic 
errors from the effective area can reach about 5\%. Taking all the statistical 
and systematical errors into account, we report the Fermi-LAT fluxes of the two
sky regions in Table \ref{tab:Fermi_flux_mask}. For reference, we also give the 
Fermi-LAT diffuse $\gamma$-ray fluxes in these sky regions without sources masks 
but just subtracting the source contribution according to Fermi-LAT source catalog. 
The non-mask fluxes and the comparison with the masked fluxes as well as the model 
predictions are presented in Appendix A. In the inner region, the masking reduces 
the resulting fluxes by a noticeable amount.

\begin{table}[!htb]
\centering
\caption{Fluxes with 1$\sigma$ uncertainties of the Galactic diffuse emission in the inner 
and outer Galaxy regions measured by Fermi-LAT. Same masks as the LHAASO analysis are adopted.}
\begin{tabular}{cccc}
\hline\hline
$\log(E/$GeV) &  $E$ & $\phi_{\rm inner}\pm\sigma_{\rm inner}$  &    $\phi_{\rm outer}\pm\sigma_{\rm outer}$        \\
                        & (GeV)              & (GeV$^{-1}$cm$^{-2}$s$^{-1}$sr$^{-1}$) & (GeV$^{-1}$cm$^{-2}$s$^{-1}$sr$^{-1}$) \\ \hline
0.00-0.18                 & 1.23              & $(1.38 \pm 0.15 ) \times 10^{-5}$    &   $(8.86 \pm 1.01 ) \times 10^{-6}$         \\
0.18-0.36                 & 1.86               & $(5.09 \pm 0.54 ) \times 10^{-6}$    &    $(3.24 \pm 0.35 ) \times 10^{-6}$        \\
0.36-0.54                 & 2.82               & $(1.86 \pm 0.20 ) \times 10^{-6}$     &   $(1.16 \pm 0.13 ) \times 10^{-6}$        \\
0.54-0.72                 & 4.26               & $(6.54 \pm 0.72 ) \times 10^{-7}$     &   $(4.00 \pm 0.45 ) \times 10^{-7}$        \\
0.72-0.90                 & 6.45               & $(2.17 \pm 0.25 ) \times 10^{-7}$      &  $(1.30 \pm 0.15 ) \times 10^{-7}$        \\
0.90-1.08                 & 9.76             & $(7.26 \pm 0.83) \times 10^{-8}$       &   $(4.10 \pm 0.48 ) \times 10^{-8}$       \\
1.08-1.26                 & 14.78             & $(2.43 \pm 0.28) \times 10^{-8}$       &    $(1.33 \pm 0.16 ) \times 10^{-8}$   \\
1.26-1.44                 & 22.36              & $(8.34  \pm 0.98) \times 10^{-9}$       &   $(4.25 \pm 0.51 ) \times 10^{-9}$   \\
1.44-1.62                 & 33.84              & $(2.86 \pm 0.32) \times 10^{-9}$       &    $(1.42 \pm 0.16 ) \times 10^{-9}$   \\
1.62-1.80                 & 51.21              & $(1.01 \pm 0.11) \times10^{-9}$        &    $(4.88 \pm 0.56 ) \times 10^{-10}$   \\ 
1.80-1.98                 & 77.50              & $(3.46\pm 0.39) \times10^{-10}$        &    $(1.61 \pm 0.20 ) \times 10^{-10}$    \\ 
1.98-2.16                 & 117.28              & $(1.26 \pm 0.15) \times10^{-10}$        &    $(5.76 \pm 0.77 ) \times 10^{-11}$    \\ 
2.16-2.34                 & 177.48              & $(4.62  \pm 0.58) \times10^{-11}$        &    $(2.09 \pm 0.31 ) \times 10^{-11}$   \\ 
2.34-2.52                 & 268.58              & $(1.48  \pm 0.22) \times10^{-11}$        &   $(4.61 \pm 1.08 ) \times 10^{-12}$     \\ 
2.52-2.70                 & 406.45             & $(5.95  \pm 1.02) \times10^{-12}$       &    $(2.67 \pm 0.63 ) \times 10^{-12}$ \\ \hline
\end{tabular}
\label{tab:Fermi_flux_mask}
\end{table}

\section{Model of diffuse emission}

\subsection{Cosmic ray propagation model}

 We use the GALPROP package\footnote{https://galprop.stanford.edu} \citep{1998ApJ...509..212S,1998ApJ...493..694M} to calculate the propagation 
of CRs and the production of secondary particles and emission. GALPROP solves 
the propagation equations numerically, with inputs of the ISM, the magnetic and 
radiation fields from astronomical observations and cross sections from nuclear 
and particle physics experiments. The three-dimensional ISM distribution is 
obtained based on surveys of HI and CO emission, together with a modeling of the 
ionized gas \citep{1976ApJ...208..346G,1988ApJ...324..248B,1991Natur.354..121C}.
The ISRF distribution is calculated based on the stellar population and dust 
absorption and re-emission \citep{2005ICRC....4...77P}. The magnetic field is assumed 
to be an analytical form of $B=5\exp(-|z|/2\,{\rm kpc}-r/10\,{\rm kpc})~\mu G$
\citep[adjusted from][]{1998ApJ...493..694M}. More details can be found in the 
manual of GALPROP\footnote{https://galprop.stanford.edu/code.php?option=manual}.

In this work we employ a one-zone cylindrically symmetric geometry to describe 
the propagation volume of CRs in the Milky Way. The radial extension of the 
propagation cylinder is fixed to be $r_{\rm max}=20$ kpc, and the halo height 
$z_h$ is a free parameter to be fitted using the CR data. The propagation of 
charged particles in the random magnetic field is characterized by a diffusion 
process, including possible convection, reacceleration by randomly moving 
magnetized plasma waves, energy losses and fragmentations due to interactions 
with the ISM \citep{1964ocr..book.....G,2007ARNPS..57..285S}.

The diffusion coefficient is parameterized as 
$D(R)=\beta^{\eta}D_0(R/4~{\rm GV})^{\delta}$, where $R$ is particle's rigidity,
$\beta$ is dimensionless velocity in unit of light speed, $\delta$ is the slope
of rigidity dependence, and $\eta$ is a phenomenological parameter introduced
to better fit the low energy secondary-to-primary ratios. The convection velocity
is assumed to be perpendicular to the Galactic plane, with magnitude being 
proportional to $z$ coordinate. The convection velocity is assumed to be 0
at $z=0$, and the gradient $dV/dz$ is a free parameter. The reacceleration is
described by a diffusion in the momentum space, whose strength is characterized 
by the Alfvenic speed $v_A$ of magnetized disturbances \citep{1994ApJ...431..705S}.
The parameters $D_0$, $\delta$, $z_h$, $dV/dz$ or $v_A$, and $\eta$ are 
the free parameters of the propagation model.

In this work we assume that the complicated spectral structures are due to 
the source injection. Alternative models including the change of the diffusion 
coefficient\footnote{For example, breaks of the secondary-to-primary ratios 
revealed by recent measurements \citep{2021PhR...894....1A,2022SciBu..67.2162D} 
suggest possible changes of the standard propagation framework of CRs
\citep{2023FrPhy..1844301M}.} \citep[either the spectrum or the spatial distribution;]
[]{2012ApJ...752...68V,2017PhRvL.119x1101G,2012ApJ...752L..13T,2018PhRvD..97f3008G,
2021PhRvD.104l3001Z} or the inclusion of discrete sources 
\citep{2013APh....50...33S,2019JCAP...10..010L,2021JCAP...05..012Z} 
are not discussed here. We employ multiple smooth breaks 
to describe those features. An exponential cutoff with characteristic cutoff 
rigidity $R_c$ is adopted to describe the knee of those particles. The injection 
spectrum takes the form of 
\begin{equation}
q_{\rm inj}=q_0 R^{-\nu_0}\exp\left(-\frac{R}{R_c}\right)\prod_{i=1}^{n}
\left[1+\left(\frac{R}{R_i}\right)^{\zeta}\right]^{\frac{\nu_{i-1}-\nu_i}{\zeta}}.
\label{eq:inj}
\end{equation}
We fix the smoothness parameter $\zeta=5$, and choose different values of $n$ for
different particles. For protons and helium, we choose $n=4$.

For electrons and positrons, we adopt the three-component model, i.e., primary 
electrons from CR acceleration sources, secondary electrons and positrons from 
CR collision with the ISM, and additional positrons and electrons from some new 
source population such as pulsars to account for the positron (and electron) 
excesses \citep{2018SCPMA..61j1002Y}. The injection spectrum of primary electrons 
is parameterized as Eq. (\ref{eq:inj}), but with $n=3$. For the additional 
positron and electron component, we use $n=2$ but with an additional low-energy 
exponential cutoff, and assume a charge symmetry between positrons and electrons. 
We further multiply a constant factor $c_{e^{\pm}}$ on the secondary positron 
and electron spectra in order to better match the low-energy data. 
Physically the $c_{e^{\pm}}$ factor may represents potential uncertainties from 
the inelastic hadronic interaction models or differences of the propagation 
between leptons and nuclei. 
The spatial distribution of primary species is assumed to follow that of supernova 
remnants or pulsars, $f(r,z)=(r/r_\odot)^{1.25}\exp\left[-3.56(r-r_\odot)/r_\odot
\right]\,\exp\left(-|z|/z_s\right)$, with $r_\odot=8.5$~kpc and $z_s=0.2$~kpc
\citep{2011ApJ...729..106T}.

\subsection{Cosmic ray fitting results}

In this work we use the CR propagation parameters obtained in 
\citet{2020JCAP...11..027Y} which fitted the fluxes of Li, Be, B, C, and O by 
AMS-02 \citep{2017PhRvL.119y1101A,2018PhRvL.120b1101A}, 
Voyager-1 \citep{2016ApJ...831...18C}, and 
ACE\footnote{http://www.srl.caltech.edu/ACE/ASC/level2/lvl2DATA\_CRIS.html}
\citep{2019SCPMA..6249511Y}. To link the Voyager-1 measurements which were expected 
to take place outside the solar system with the measurements on top of the Earth, 
a force-field solar modulation model \citep{1968ApJ...154.1011G} is used.
Two typical setups of the CR propagation model, the diffusion + convection (DC) 
one\footnote{However, as shown in \citet{2020JCAP...11..027Y}, the fitted convection 
velocity is close to 0 and the model reduces to a plain diffusion model.} 
and the diffusion + reacceleration (DR) one are assumed. 
The improvement of the fitting for a model including both reacceleration and 
convection is expected to be limited since a large convection velocity is disfavored 
by the data and the DR model already fits the data well \citep{2020JCAP...11..027Y}.
The best-fitting propagation parameters are given in Table \ref{tab:prop} 
\citep{2020JCAP...11..027Y}, and the comparisons between the best-fitting model 
predictions and observational data are shown in Fig.~\ref{fig:nuclei} of Appendix B. 
More details about the fitting procedure and results can be found in 
\citet{2020JCAP...11..027Y}.

\begin{table}[!htb]
\centering
\small
\caption{Propagation parameters \citep{2020JCAP...11..027Y}.}
\begin{tabular}{ccccccc}
    \hline\hline
    Model & $D_0$ & $\delta$ & $z_h$ & $dV/dz$ & $v_A$ & $\eta$ \\
    & ($10^{28}$~cm$^2$~s$^{-1}$) & & (kpc) & (km~s$^{-1}$~kpc$^{-1}$) & (km~s$^{-1}$) & \\
    \hline
    DC  & $4.10\pm0.34$ & $0.477\pm0.003$ & $4.9\pm0.5$ & $<0.86$ (95\% C.L.) & --- & $-1.51\pm0.03$ \\
    DR  & $7.69\pm0.60$ & $0.362\pm0.004$ & $6.3\pm0.7$ & ---   & $33.76\pm0.67$ & $-0.05\pm0.04$ \\
    \hline
\end{tabular}
\label{tab:prop}
\end{table}

\begin{table}[!htb]
\centering
\small
\caption{Source injection and solar modulation parameters of p and He nuclei.}
\begin{tabular}{c|cccc|cccc}
\hline\hline
   & \multicolumn{4}{c|}{Proton} & \multicolumn{4}{c}{Helium} \\
 & DC-high & DC-low & DR-high & DR-low & DC-high & DC-low & DR-high & DR-low \\ \hline
$\nu_0$ &$1.70$&$1.70$&$2.06$&$2.06$ & $1.60$&$1.60$&$1.46$&$1.46$ \\
$\nu_1$ &$2.41$&$2.41$&$2.43$&$2.43$ & $2.33$&$2.33$&$2.36$&$2.36$ \\
$\nu_2$ &$2.12$&$2.12$&$2.22$&$2.22$ & $2.02$&$2.02$&$2.12$&$2.12$ \\
$\nu_3$ &$2.42$&$2.42$&$2.52$&$2.52$ & $2.32$&$2.32$&$2.42$&$2.42$ \\
$\nu_4$ &$2.08$&$2.22$&$2.18$&$2.32$ & $1.98$&$2.18$&$2.08$&$2.28$ \\
$R_1/{\rm GV}$ &$1.69$&$1.69$&$13.9$&$13.9$ & $1.41$&$1.41$&$1.99$&$1.99$ \\
$R_2/{\rm TV}$ &$0.31$&$0.31$&$0.50$&$0.50$ & $0.50$&$0.50$&$0.65$&$0.65$ \\
$R_3/{\rm TV}$ &$15.0$&$15.0$&$15.0$&$15.0$ & $15.0$&$15.0$&$15.0$&$15.0$ \\
$R_4/{\rm TV}$ &$100.0$&$100.0$&$100.0$&$100.0$ & $100.0$&$100.0$&$100.0$&$100.0$ \\
$R_c/{\rm PV}$ &$12.0$&$4.0$&$12.0$&$4.0$ & $6.0$&$4.0$&$6.0$&$4.0$ \\
$\Phi/{\rm GV}$&$0.685$&$0.685$&$0.600$&$0.600$ & $0.600$&$0.600$&$0.700$&$0.700$ \\
\hline
\end{tabular}
\label{tab:inj_phe}
\end{table}

\begin{figure*}[!htb]
\centering
\includegraphics[width=0.48\textwidth]{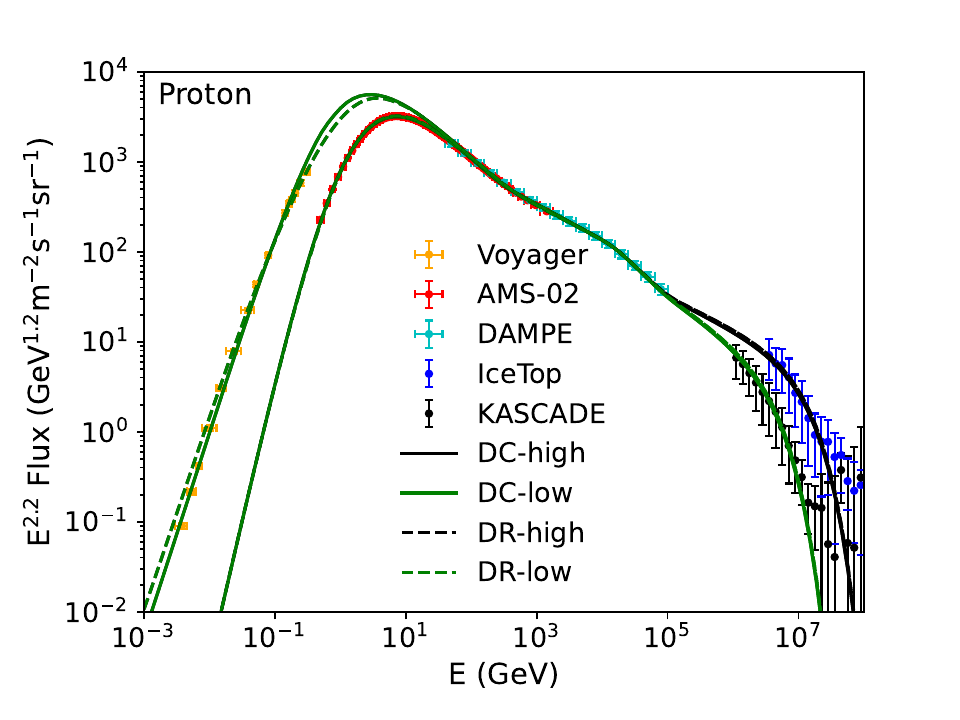}
\includegraphics[width=0.48\textwidth]{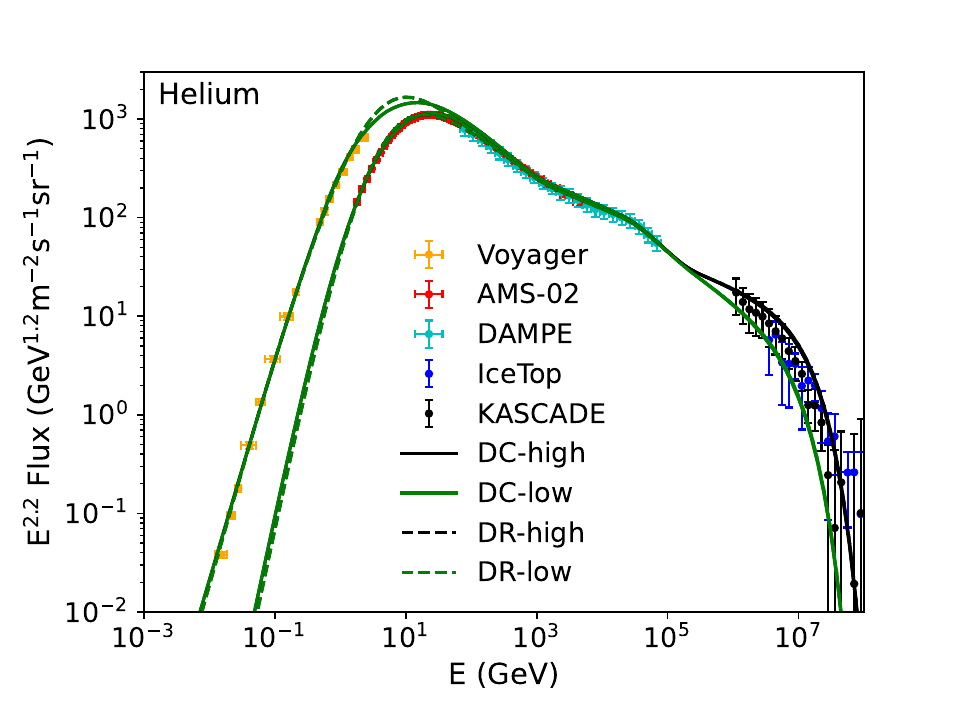}
\includegraphics[width=0.48\textwidth]{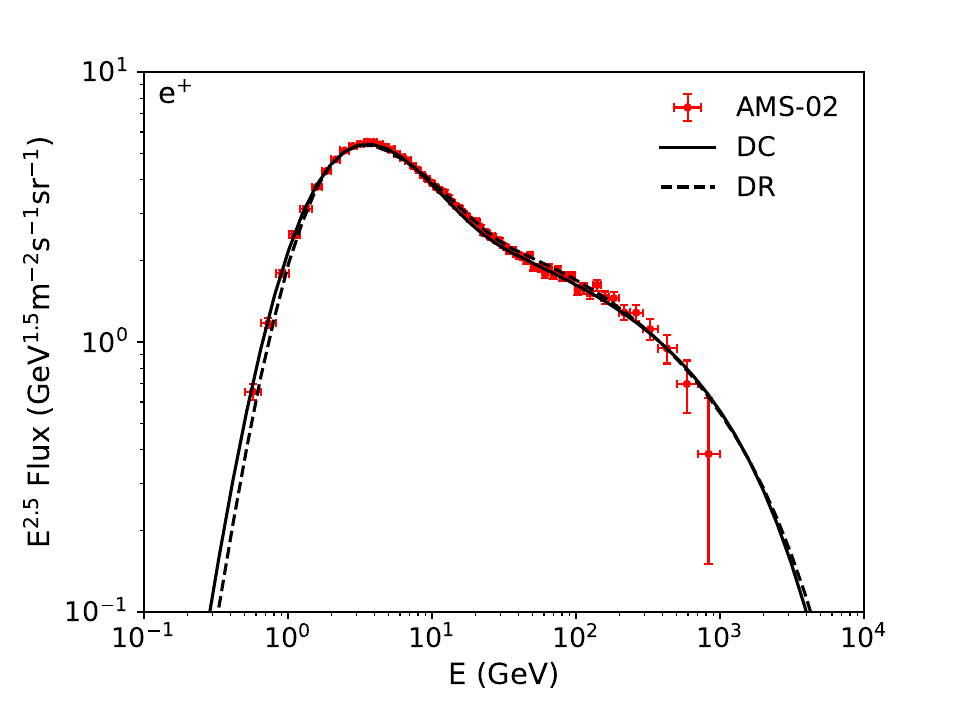}
\includegraphics[width=0.48\textwidth]{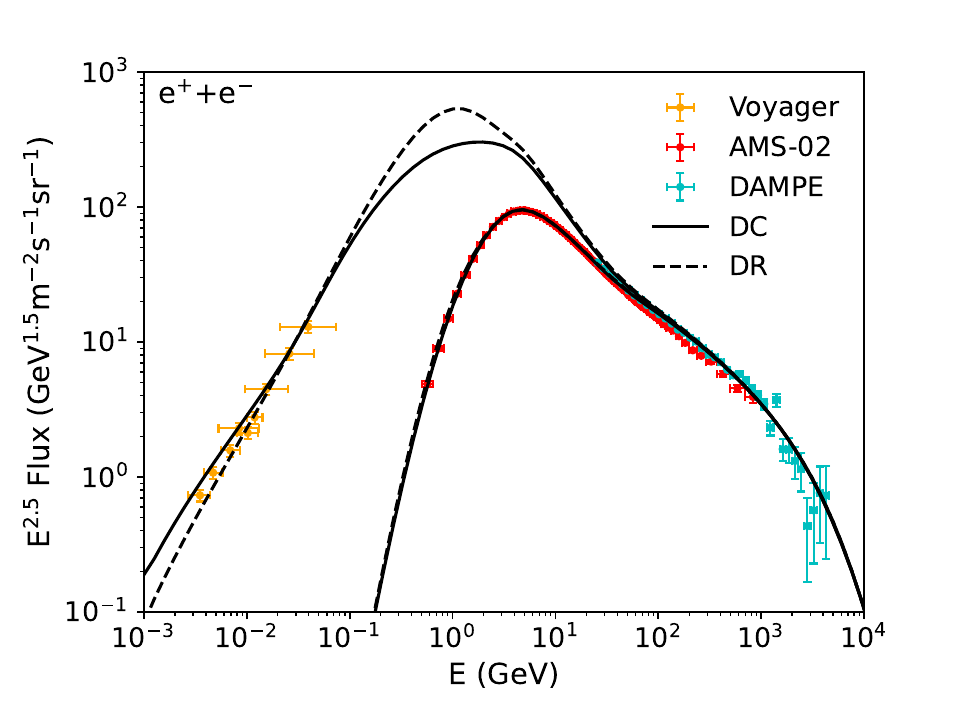}
\caption{Model predicted spectra of protons (top-left), helium nuclei (top-right), 
positrons (bottom-left), and total electrons plus positrons (bottom-right),
compared with the measurements. Solid lines are for the DC model, and dashed
lines are for the DR model. For protons and helium nuclei, the ``-high'' and
``-low'' results correspond to the uncertainties of the measurements by indirect
experiments. Higher curves at low energies ($\lesssim10$ GeV) are unmodulated
spectra to fit the Voyager-1 data.}
\label{fig:phe_ep}
\end{figure*}

Fixing the propagation parameters derived from fitting to the intermediate 
mass nuclei, we get the injection spectra of protons, helium nuclei, electrons, 
and positrons which are most relevant to the calculation of diffuse $\gamma$-rays.
The proton and helium data used include those measured by Voyager-1 
\citep{2016ApJ...831...18C}, AMS-02 \citep{2021PhR...894....1A,2017PhRvL.119y1101A}, 
DAMPE \citep{2019SciA....5.3793A,2021PhRvL.126t1102A}, 
IceTop \citep{2019PhRvD.100h2002A} and KASCADE-Grande \citep{2013APh....47...54A}. 
Since the measurements around the knee region show remarkable differences among 
different experiments, we try to match the data with high and low flux assumptions 
to give an uncertainty band of the results. For positrons and electrons we use 
the positron spectrum measured by AMS-02 \citep{2019PhRvL.122d1102A}, the total 
electron plus positron spectra measured by Voyager-1 \citep{2016ApJ...831...18C}, 
AMS-02 \citep{2019PhRvL.122j1101A}, and DAMPE \citep{2017Natur.552...63D}. 
The derived source parameters are given in Tables \ref{tab:inj_phe} and 
\ref{tab:inj_ep}. Fig.~\ref{fig:phe_ep} shows the spectra of protons, helium 
nuclei, positrons, and total electrons plus positrons, compared with the data. 
The major spectral structures of these particles can be properly reproduced.

\begin{table}[!htb]
\centering
\small
\caption{Source injection and solar modulation parameters of electrons and positrons.}
\begin{tabular}{cccccccccc}
\hline\hline
Model  & $\nu_0^-$ & $\nu_1^-$ & $\nu_2^-$ & $\nu_3^-$ & $R_1^-$/GV & $R_2^-$/GV & $R_3^-$/GV &$R_c^-$/TV & $\Phi^-$/GV \\
DC & $2.53$ & $1.68$ & $2.92$ & $2.41$ & $0.044$ & $4.70$ & $49.1$ & $6.68$ & 0.876\\
DR & $2.33$ & $0.01$ & $2.88$ & $2.45$ & $0.950$ & $4.19$ & $55.7$ & $6.27$ & 1.000\\
\hline
Model & $c_{e^{\pm}}$ & $\nu_1^+$ & $\nu_2^+$ & $R_1^+$/GV & $R_{c,{\rm low}}^+$/GV & $R_{c,{\rm high}}^+$/TV & 
$A_{\rm new}^{\dagger}$ & $\Phi^+$/GV \\
DC & $1.30$ & $3.04$ & $1.88$ & $24.6$ & $0.97$ & $2.45$ & $8.0\times10^{-30}$ & 0.783 \\
DR & $1.00$ & $3.04$ & $2.08$ & $31.2$ & $4.89$ & $3.42$ & $1.8\times10^{-29}$ & 0.950 \\
\hline
\end{tabular}\\
\label{tab:inj_ep}
$^\dagger$Pre-propagated normalization of new source population at 1~GeV in unit of cm$^{-3}$~s$^{-1}$~MeV$^{-1}$.
\end{table}

\subsection{Calculation of diffuse $\gamma$-ray emission}

Using the CR model parameters we calculate the diffuse $\gamma$-ray emission
from the neutral pion decay, the bremsstrahlung and inverse Compton scattering
components with GALPROP. The gamma-ray production from secondary particles by
inelastic hadronic interactions is calculated with the {\tt AAfrag} package 
\citep{2019CoPhC.24506846K}, which uses the updated QGSJET-II-04m model for
hadronic interactions with parameters tuned based on recent data from the LHCf, 
LHCb, and NA61 experiments.

For VHE $\gamma$-ray photons, the pair production attenuation due to 
scattering with the ISRF background is important, which needs to be included. 
The optical depth of a high-energy photon with energy $E_{\gamma}$ and 
location ${\boldsymbol x}=(R,z,\theta)$ is
\begin{equation}
\tau(E_{\gamma},{\boldsymbol x})=\int_{l.o.s.}dl\int\frac{1-\mu}{2}d\mu
\int \frac{dn}{d\epsilon} \sigma_{\gamma\gamma\to e^+e^-}(s)d\epsilon,
\end{equation}
where $\mu=\cos\theta'$ with $\theta'$ being the angle between the momenta
of the incoming photon and the background photon, and $dn/d\epsilon$ is 
the number density of the background radiation field. The pair production
cross section is \citep{2006A&A...449..641Z}
\begin{equation}
\sigma_{\gamma\gamma\to e^+e^-}(s)=\sigma_{\rm T}\cdot\frac{3m_e^2}{2s}
\cdot\left[-\frac{p}{E}\left(1+\frac{4m_e^2}{s}\right)+\left(1+
\frac{4m_e^2}{s}\left(1-\frac{2m_e^2}{s}\right)\right)
\ln\frac{(E+p)^2}{m_e^2}\right],
\end{equation}
where $\sigma_T$ is the Thomson cross section, $E\equiv\sqrt{s}/2$ and 
$p\equiv\sqrt{E^2-m_e^2}$ are the energy and momentum of the electron or
positron in the center-of-momentum system, $s=2E_{\gamma}\epsilon(1-\mu)$ 
is the center-of-momentum energy.
An attenuation factor $e^{-\tau}$ has been multiplied to the calculated 
$\gamma$-ray emissivity before the line-of-sight integration.

\section{Confronting model predictions with measurements}

\begin{figure*}[!htb]
\centering
\includegraphics[width=0.48\textwidth]{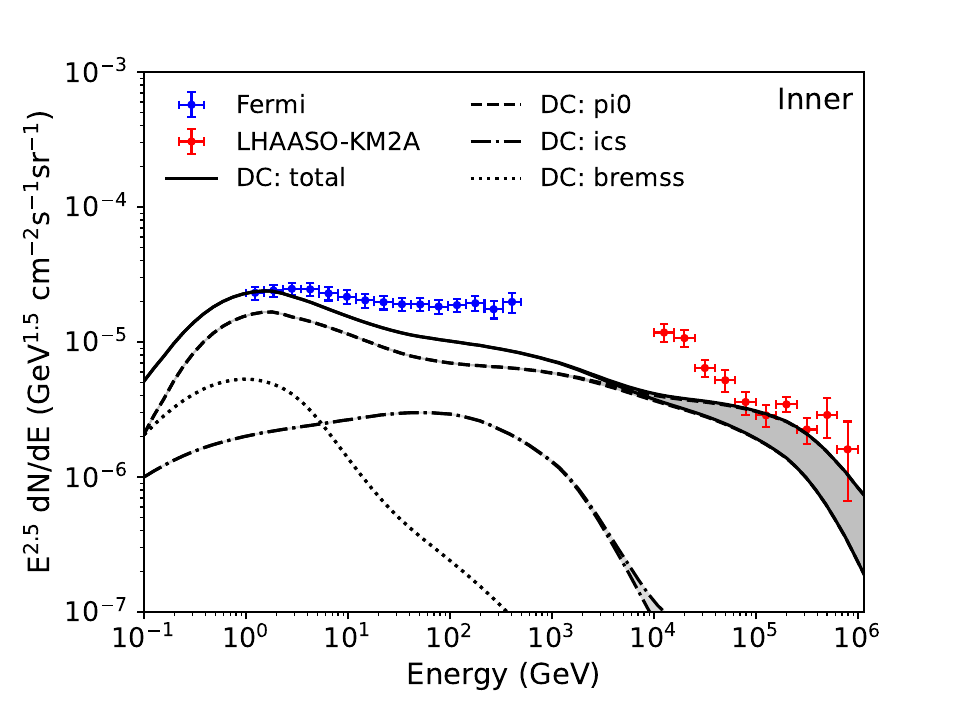}
\includegraphics[width=0.48\textwidth]{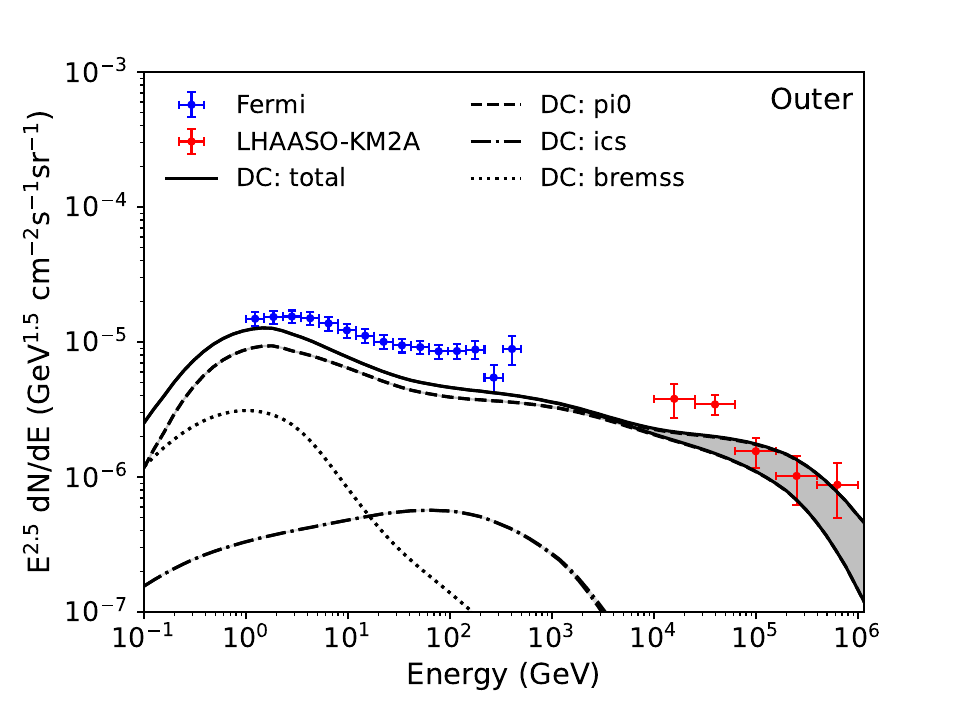}
\includegraphics[width=0.48\textwidth]{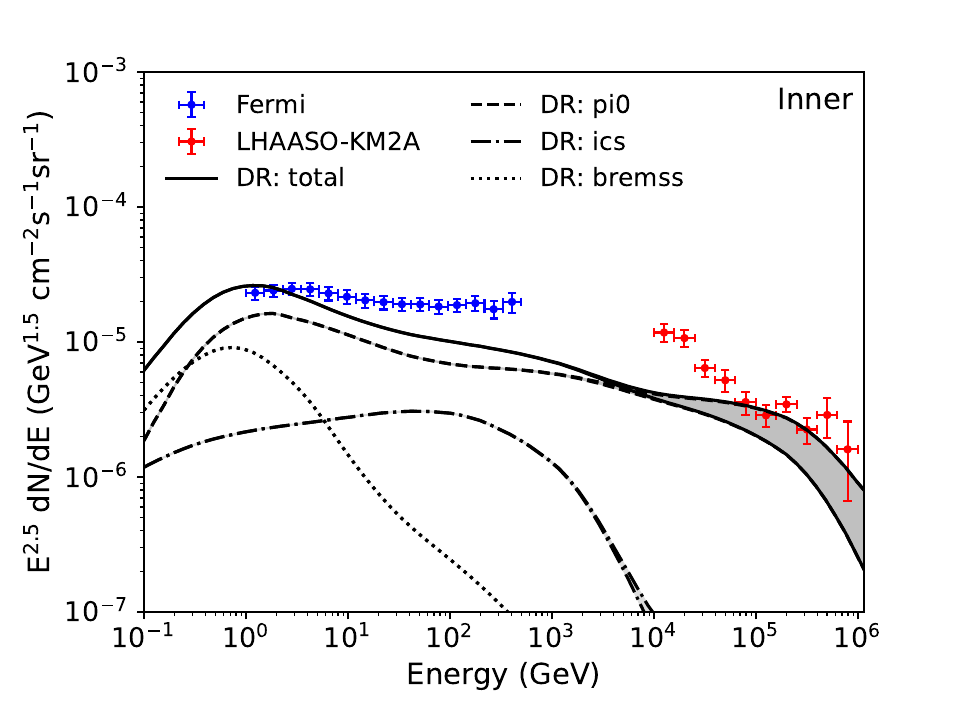}
\includegraphics[width=0.48\textwidth]{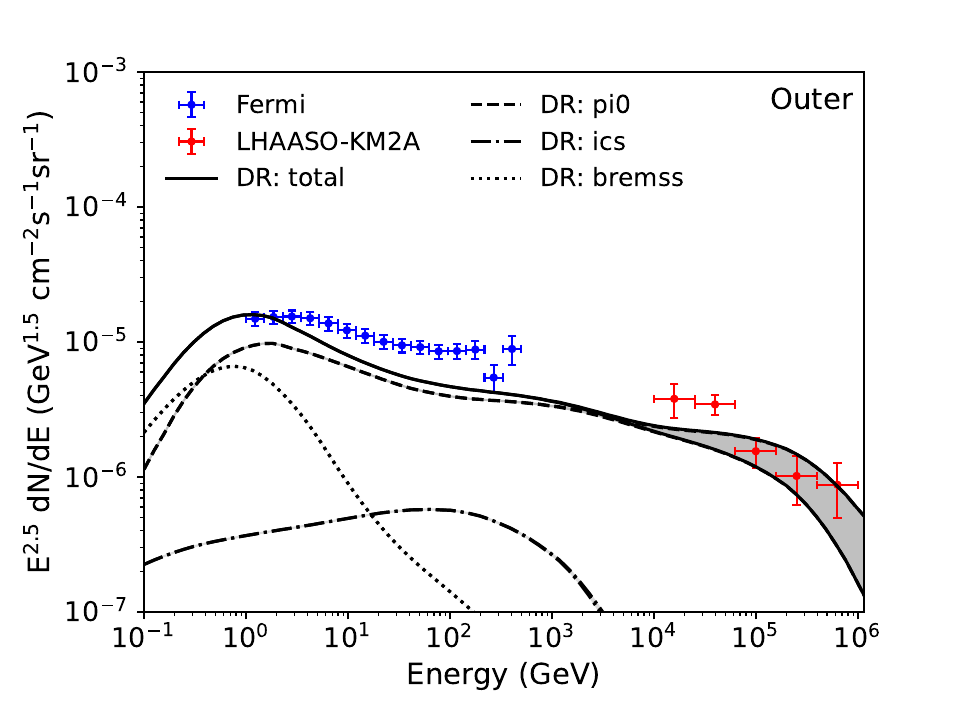}
\caption{Wide-band spectra of the diffuse $\gamma$-ray emission. Left panels are
for the inner Galaxy region, and right panels are for the outer Galaxy region.
Top two panels show the model predictions of the DC model, and bottom two are for 
the DR model.}
\label{fig:diffuse}
\end{figure*}

The model predicted diffuse $\gamma$-ray spectra, together with the measurements
by LHAASO and Fermi-LAT are shown in Fig.~\ref{fig:diffuse}. For all the model
results, we apply the same masks as shown in \citet{LHAASO-diffuse-2023} to enable 
a self-consistent comparison. We can see that the model predictions are only 
consistent with the data at the lowest energies (less than a few GeV).
Clear excesses of the emission from a few GeV to $\sim60$ TeV are visible.
At the highest energy end (above $\sim 60$ TeV), the upper edges of the 
predicted results are marginally consistent with the data. 
To see the effect of source masks, we also show the comparison between 
the DC model predictions with the Fermi-LAT data without source masks in 
Fig.~\ref{fig:nomask_model} in Appendix A. Similar excesses above a few GeV 
can also be seen clearly.

\begin{figure*}[!htb]
\centering
\includegraphics[width=0.48\textwidth]{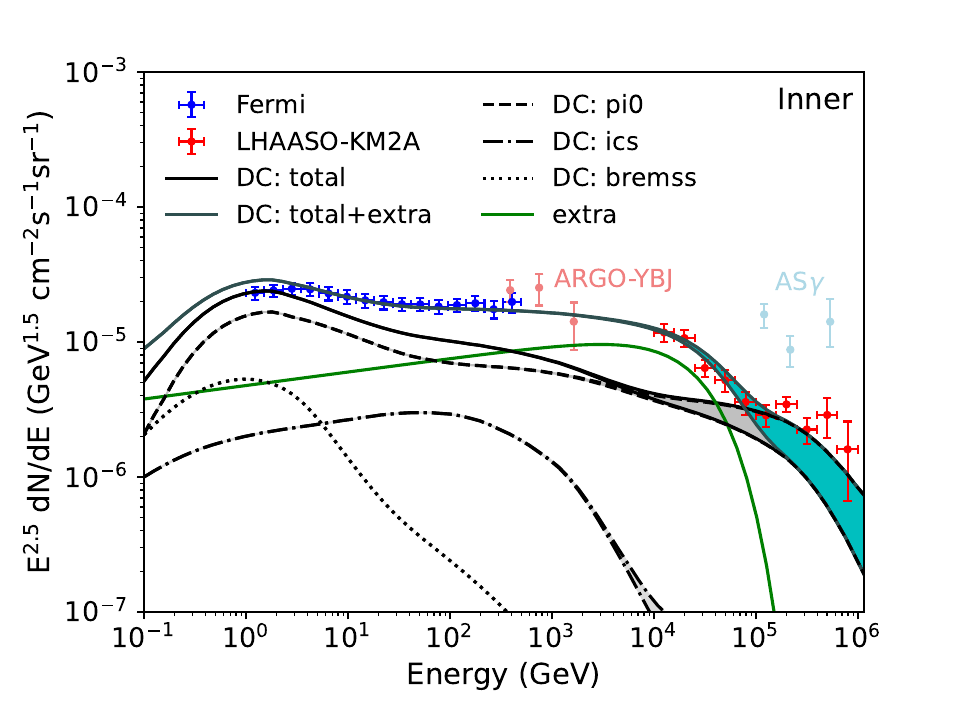}
\includegraphics[width=0.48\textwidth]{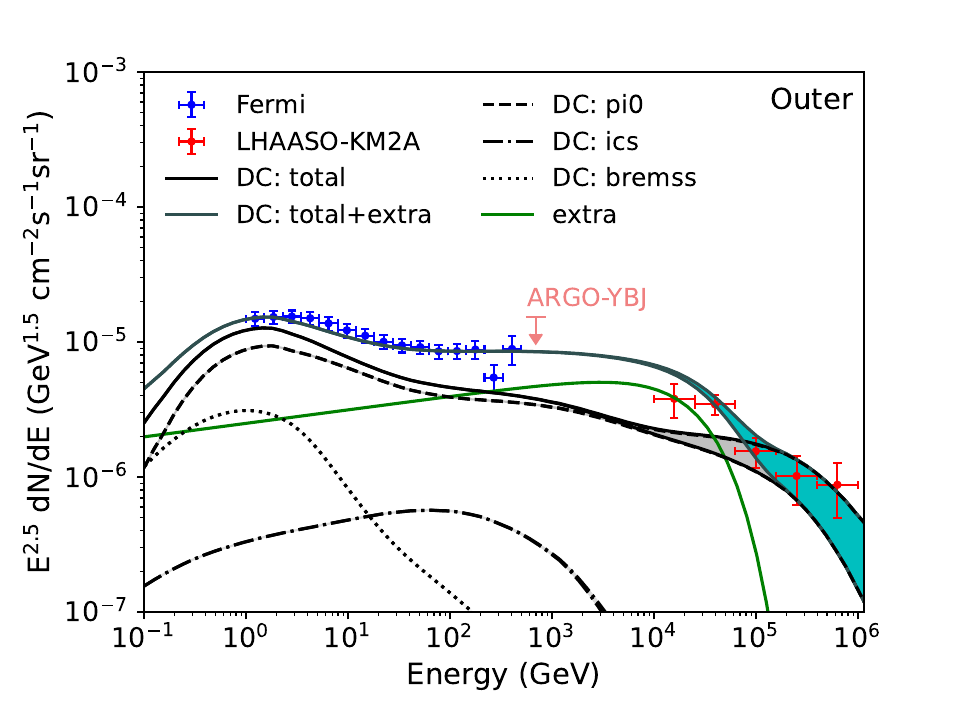}
\caption{Wide-band spectra of the diffuse $\gamma$-ray emission for the DC
propagation model, assuming an extra component with an ECPL spectrum. The left panel
is for the inner Galaxy region, and the right one is for the outer Galaxy region.
Note that the ARGO-YBJ and AS$\gamma$ data in the left panel are for a different 
region of $25^\circ<l<100^\circ$ and $|b|<5^\circ$, and the ARGO-YBJ upper limit 
for the right panel is for $130^\circ<l<200^\circ$ and $|b|<5^\circ$
\citep{2015ApJ...806...20B,2021PhRvL.126n1101A}. Source masks of these analyses
are also different.}
\label{fig:diffuse-DCextra}
\end{figure*}

\begin{figure*}[!htb]
\centering
\includegraphics[width=0.48\textwidth]{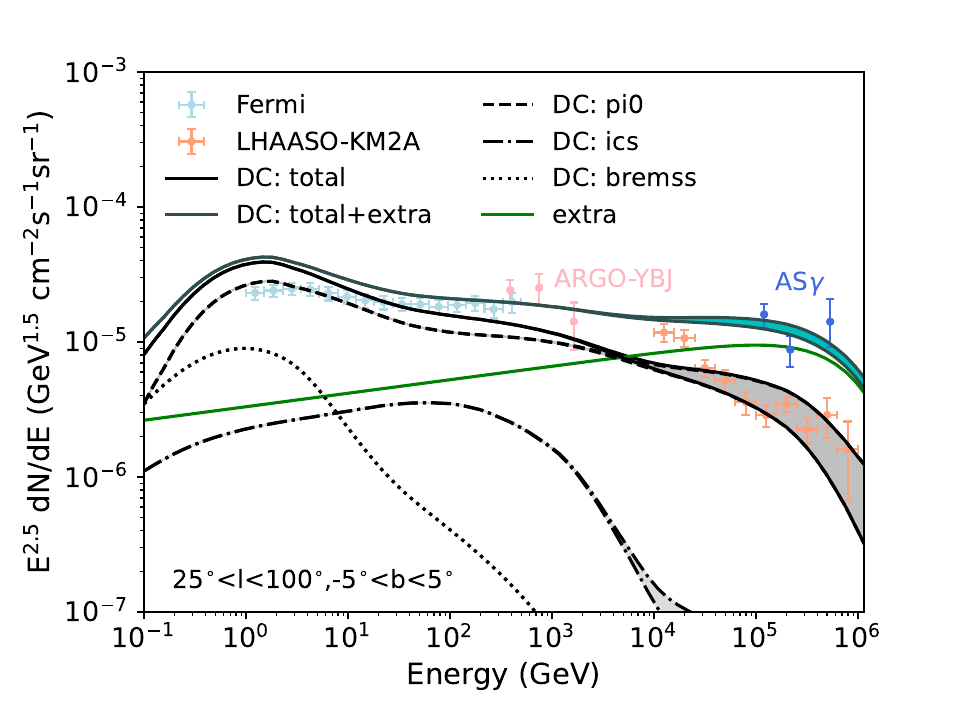}
\caption{Comparison between the DC model predicted diffuse $\gamma$-ray spectra and the 
AS$\gamma$ data \citep{2021PhRvL.126n1101A}, for a region of $25^\circ<l<100^\circ$ and 
$|b|<5^\circ$. An extra ECPL component is also added to account for the data. The same mask 
as adopted in the AS$\gamma$ analysis is applied to the model curves. The other data are the 
same as Fig.~\ref{fig:diffuse-DCextra}, whose ROIs are different from the ROI of AS$\gamma$.}
\label{fig:diffuse-DCmodel-asgammaMASK}
\end{figure*}

Excesses from the Galactic plane for energies above a few GeV were also revealed 
through comparing the measurements with the models \citep{2007APh....27...10P,2012ApJ...750....3A}.
One possible origin of those excesses is undetected point-like sources. A widely 
discussed candidate population of unresolved sources is pulsars or pulsar wind nebulae 
\citep{2000A&A...362..937A,2018PhRvL.120l1101L,2022ApJ...928...19V,2023arXiv230412574Y}. 
To see whether such an additional component can fit the data, we simply add an
exponential-cutoff-power-law (ECPL) function to the total emission for the DC model, 
with a spectral index of $-2.40$ and cutoff energy of $30$ TeV. The results can match
the data from GeV to PeV well, as shown in Fig.~\ref{fig:diffuse-DCextra}. 
Slight excesses in the inner region for $E>100$ TeV may indicate that unresolved 
PeVatrons still exist in the data. 
The normalization of the ECPL component in the inner region is about two times
of that in the outer region, which is very close to the expected ratio according to the 
source distribution $f(r,z)$ as described in Sec. 3.1 (${\rm inner/outer}=1.85$). 
It indicates that they might be from a single population with spatial distribution 
close to the supernova remnants or pulsars.
We also note that the longitudinal distribution of the observed diffuse emission 
seems not closely follow the gas distribution \citep{LHAASO-diffuse-2023}, 
which may further support the unresolved source origin of the excesses.

Other scenarios to give higher diffuse fluxes may include hadronic interactions 
between freshly accelerated CRs and the medium surrounding acceleration sources
\citep{2019PhRvD.100f3020Y,2022PhRvD.105b3002Z}, the spatial variations of the 
CR spectra \citep{2016PhRvD..93l3007Y,2018PhRvD..98d3003L,2023A&A...672A..58D},
or the transition of different propagation regimes \citep{2022SNAS....4...15R}. 
However, since the excesses mainly exist in an energy range from several GeV to 
60 TeV, the required spectral shape of the excess component is distinct from that 
of a continuous CR spectrum like the background sea. Whether these models can 
account for the data needs detailed studies.

In Fig.~\ref{fig:diffuse-DCmodel-asgammaMASK} we compare the model predictions
with the AS$\gamma$ data for sky region of $25^\circ<l<100^\circ$ and $|b|<5^\circ$,
with the mask adopted in the AS$\gamma$ analysis \citep{2021PhRvL.126n1101A}.
One can see that the model predictions are significantly lower than the data.
Adding an extra ECPL component with index of $-2.40$ and cutoff energy of 1 PeV
may account for the data. However, it should be noted that the mask region of AS$\gamma$, 
i.e., $0.5^{\circ}$ radius disks around detected sources, might not be enough to 
exclude the emission from those sources \citep{2023arXiv230517030C}.

\section{Conclusion and discussion}

Motivated by the recent measurements of diffuse $\gamma$-ray emission from the
Galactic plane by LHAASO \citep{LHAASO-diffuse-2023}, we study the emission 
model from interactions between Galactic CRs and the ISM, in a wide energy range 
from GeV to PeV. We re-analyze the Fermi-LAT data to derive the diffuse emission
from 1 to 500 GeV in the two sky regions consistent with those adopted by LHAASO.
This will allow us to have a self-consistent wide-band comparison between model 
and data. Using the latest measurements of local CR spectra, we get improved 
constraints on the propagation and injection parameters of CRs under the framework 
of one-zone, homogeneous, and isotropic diffusion scenario. Possible reacceleration 
or convection transportation of CRs is included in the model. Based on the CR 
parameters, we obtain the diffuse $\gamma$-ray emission, and compare the model 
predictions with the Fermi-LAT and LHAASO data. We find that the model under-predicts
the diffuse fluxes between several GeV and $\sim60$ TeV. Such excesses may be
explained by an extra component characterized by an ECPL spectrum
($\propto E^{-2.40}\exp{(-E/30~{\rm TeV})}$). 

In the current LHAASO analysis, a considerable fraction of the sky has been
masked to reduce the impact from detected sources. While the diffuse emission is
expected to be the brightest toward the central plane of the Galaxy, such masks
remove a significant part of the diffuse emission. Refined treatment of the source 
removal will be helpful in better constraining the modelling of diffuse $\gamma$
rays and Galactic CRs. Improved measurements of the variations of energy spectra 
across the Galactic plane will also be very important.

The propagation model of CRs is under a framework of homogeneous and 
isotropic diffusion, which may be over-simplified. More sophisticated models with 
spatially-dependent propagation as inferred from observations of pulsar halos
\citep{2017Sci...358..911A,2021PhRvL.126x1103A} or anisotropic diffusion along 
large-scale magnetic fields \citep{2023arXiv230510251G} may give different 
predictions of the diffuse $\gamma$-ray spectrum. Furthermore, local sources may 
also contribute to the observed CR spectra
\citep{2019JCAP...10..010L,2021JCAP...05..012Z}, resulting in different CR spectra
throughout the Galaxy. Detailed comparison between these alternative models and the 
data will be interesting for future works.

We note that the direct measurements of CR spectra around PeV have large 
uncertainties. New measurements of the proton and helium spectra across PeV 
energies by LHAASO are expected to be very useful in reducing the uncertainties 
of local CR spectra. Furthermore, the $\gamma$-ray production cross section of 
inelastic hadronic interactions also experience large uncertainties 
\citep{2006ApJ...647..692K,2014PhRvD..90l3014K,2021PhRvD.104l3027K}.
Additional uncertainties may come from the gas distribution used in the model
\citep[e.g.,][]{2017A&A...601A..78R}. 
All these uncertainties prevent a precise prediction of the diffuse $\gamma$-ray 
fluxes, particularly at the highest energies. Joint efforts in improving all these 
aspects are necessary to better understand the interactions of CRs in the ISM.

The hadronic interaction would also produce diffuse neutrinos associated with
$\gamma$-rays, which have recently been detected by IceCube \citep{IceCube2023}.
The neutrino fluxes measured by IceCube, which are model-dependent, may contain
both the diffuse contribution and unresolved sources. Therefore, the comparison 
between the neutrino data and the $\gamma$-ray data should be done carefully
\citep{2023arXiv230712363Y,2023arXiv230702905F}. In any case, the multi-messenger
data should be very helpful in testing the origin of the VHE to UHE emission of 
the Galactic plane.

%Current observations of ANTARES and IceCube give only upper limits of any Galactic-disk 
%related neutrino emission \citep{2017ApJ...849...67A,2017PhRvD..96f2001A}. With the 
%increase of exposure of neutrino experiments, it is likely that the Galactic component 
%of neutrinos can be finally unveiled, which can critically address the hadronic or 
%leptonic origin of the ultra-high-energy diffuse $\gamma$-rays.

\section*{Acknowledgements}
We thank Jun Li for helpful discussion. 
This work is supported by the National Natural Science Foundation of China 
(Nos. 12220101003, 12322302), the Project for Young Scientists in Basic Research 
of Chinese Academy of Sciences (No. YSBR-061), the Chinese Academy of Sciences, 
and the Program for Innovative Talents and Entrepreneur in Jiangsu.

\bibliographystyle{apj}
\bibliography{refs}

\begin{thebibliography}{}
\expandafter\ifx\csname natexlab\endcsname\relax\def\natexlab#1{#1}\fi

\bibitem[{{Aartsen} {et~al.}(2019){Aartsen}, {Ackermann}, {Adams}, {Aguilar},
  {Ahlers}, {Ahrens}, {Alispach}, {Andeen}, {Anderson}, {Ansseau}, {Anton},
  {Arg{\"u}elles}, {Auffenberg}, {Axani}, {Backes}, {Bagherpour}, {Bai},
  {Barbano}, {Barwick}, {Baum}, {Baur}, {Bay}, {Beatty}, {Becker}, {Becker
  Tjus}, {BenZvi}, {Berley}, {Bernardini}, {Besson}, {Binder}, {Bindig},
  {Blaufuss}, {Blot}, {Bohm}, {B{\"o}rner}, {B{\"o}ser}, {Botner},
  {B{\"o}ttcher}, {Bourbeau}, {Bourbeau}, {Bradascio}, {Braun}, {Bretz},
  {Bron}, {Brostean-Kaiser}, {Burgman}, {Buscher}, {Busse}, {Carver}, {Chen},
  {Cheung}, {Chirkin}, {Clark}, {Classen}, {Collin}, {Conrad}, {Coppin},
  {Correa}, {Cowen}, {Cross}, {Dave}, {de Andr{\'e}}, {De Clercq}, {DeLaunay},
  {Dembinski}, {Deoskar}, {De Ridder}, {Desiati}, {de Vries}, {de Wasseige},
  {de With}, {DeYoung}, {Diaz}, {D{\'\i}az-V{\'e}lez}, {Dujmovic}, {Dunkman},
  {Dvorak}, {Eberhardt}, {Ehrhardt}, {Eller}, {Evenson}, {Fahey}, {Fazely},
  {Felde}, {Feusels}, {Filimonov}, {Finley}, {Franckowiak}, {Friedman},
  {Fritz}, {Gaisser}, {Gallagher}, {Ganster}, {Garrappa}, {Gerhardt},
  {Ghorbani}, {Glauch}, {Gl{\"u}senkamp}, {Goldschmidt}, {Gonzalez}, {Grant},
  {Griffith}, {G{\"u}nder}, {G{\"u}nd{\"u}z}, {Haack}, {Hallgren}, {Halve},
  {Halzen}, {Hanson}, {Hebecker}, {Heereman}, {Heix}, {Helbing}, {Hellauer},
  {Henningsen}, {Hickford}, {Hignight}, {Hill}, {Hoffman}, {Hoffmann},
  {Hoinka}, {Hokanson-Fasig}, {Hoshina}, {Huang}, {Huber}, {Hultqvist},
  {H{\"u}nnefeld}, {Hussain}, {In}, {Iovine}, {Ishihara}, {Jacobi},
  {Japaridze}, {Jeong}, {Jero}, {Jones}, {Jonske}, {Joppe}, {Kang}, {Kappes},
  {Kappesser}, {Karg}, {Karl}, {Karle}, {Katz}, {Kauer}, {Kelley},
  {Kheirandish}, {Kim}, {Kintscher}, {Kiryluk}, {Kittler}, {Klein}, {Koirala},
  {Kolanoski}, {K{\"o}pke}, {Kopper}, {Kopper}, {Koskinen}, {Kowalski},
  {Krings}, {Kr{\"u}ckl}, {Kulacz}, {Kunwar}, {Kurahashi}, {Kyriacou},
  {Labare}, {Lanfranchi}, {Larson}, {Lauber}, {Lazar}, {Leonard}, {Leuermann},
  {Liu}, {Lohfink}, {Lozano Mariscal}, {Lu}, {Lucarelli}, {L{\"u}nemann},
  {Luszczak}, {Madsen}, {Maggi}, {Mahn}, {Makino}, {Mallik}, {Mallot},
  {Mancina}, {Mari{\c{s}}}, {Maruyama}, {Mase}, {Maunu}, {Meagher}, {Medici},
  {Medina}, {Meier}, {Meighen-Berger}, {Menne}, {Merino}, {Meures}, {Miarecki},
  {Micallef}, {Moment{\'e}}, {Montaruli}, {Moore}, {Morse}, {Moulai}, {Muth},
  {Nagai}, {Nahnhauer}, {Nakarmi}, {Naumann}, {Neer}, {Niederhausen},
  {Nowicki}, {Nygren}, {Obertacke Pollmann}, {Olivas}, {O'Murchadha},
  {O'Sullivan}, {Palczewski}, {Pandya}, {Pankova}, {Park}, {Peiffer},
  {P{\'e}rez de los Heros}, {Philippen}, {Pieloth}, {Pinat}, {Pizzuto}, {Plum},
  {Porcelli}, {Price}, {Przybylski}, {Raab}, {Raissi}, {Rameez}, {Rauch},
  {Rawlins}, {Rea}, {Reimann}, {Relethford}, {Renzi}, {Resconi}, {Rhode},
  {Richman}, {Robertson}, {Rongen}, {Rott}, {Ruhe}, {Ryckbosch}, {Rysewyk},
  {Safa}, {Sanchez Herrera}, {Sandrock}, {Sandroos}, {Santander}, {Sarkar},
  {Sarkar}, {Satalecka}, {Schaufel}, {Schlunder}, {Schmidt}, {Schneider},
  {Schneider}, {Schumacher}, {Sclafani}, {Seckel}, {Seunarine}, {Shefali},
  {Silva}, {Snihur}, {Soedingrekso}, {Soldin}, {Song}, {Spiczak}, {Spiering},
  {Stachurska}, {Stamatikos}, {Stanev}, {Stasik}, {Stein}, {Stettner},
  {Steuer}, {Stezelberger}, {Stokstad}, {St{\"o}{\ss}l}, {Strotjohann},
  {St{\"u}rwald}, {Stuttard}, {Sullivan}, {Sutherland}, {Taboada}, {Tenholt},
  {Ter-Antonyan}, {Terliuk}, {Tilav}, {Tomankova}, {T{\"o}nnis}, {Toscano},
  {Tosi}, {Tselengidou}, {Tung}, {Turcati}, {Turcotte}, {Turley}, {Ty},
  {Unger}, {Unland Elorrieta}, {Usner}, {Vandenbroucke}, {Van Driessche}, {van
  Eijk}, {van Eijndhoven}, {Vanheule}, {van Santen}, {Vraeghe}, {Walck},
  {Wallace}, {Wallraff}, {Wandkowsky}, {Watson}, {Weaver}, {Weiss}, {Weldert},
  {Wendt}, {Werthebach}, {Westerhoff}, {Whelan}, {Whitehorn}, {Wiebe},
  {Wiebusch}, {Wille}, {Williams}, {Wills}, {Wolf}, {Wood}, {Wood},
  {Woschnagg}, {Wrede}, {Xu}, {Xu}, {Xu}, {Yanez}, {Yodh}, {Yoshida}, {Yuan},
  {Z{\"o}cklein}, \& {IceCube Collaboration}}]{2019PhRvD.100h2002A}
{Aartsen}, M.~G., {Ackermann}, M., {Adams}, J., {et~al.} 2019, \prd, 100,
  082002

\bibitem[{{Abdo} {et~al.}(2007){Abdo}, {Allen}, {Berley}, {Blaufuss},
  {Casanova}, {Chen}, {Coyne}, {Delay}, {Dingus}, {Ellsworth}, {Fleysher},
  {Fleysher}, {Gebauer}, {Gonzalez}, {Goodman}, {Hays}, {Hoffman}, {Kolterman},
  {Kelley}, {Lansdell}, {Linnemann}, {McEnery}, {Mincer}, {Moskalenko},
  {Nemethy}, {Noyes}, {Ryan}, {Samuelson}, {Saz Parkinson}, {Schneider},
  {Shoup}, {Sinnis}, {Smith}, {Strong}, {Sullivan}, {Vasileiou}, {Walker},
  {Williams}, {Xu}, \& {Yodh}}]{2007ApJ...658L..33A}
{Abdo}, A.~A., {Allen}, B., {Berley}, D., {et~al.} 2007, \apjl, 658, L33

\bibitem[{{Abdo} {et~al.}(2008){Abdo}, {Allen}, {Aune}, {Berley}, {Blaufuss},
  {Casanova}, {Chen}, {Dingus}, {Ellsworth}, {Fleysher}, {Fleysher},
  {Gonzalez}, {Goodman}, {Hoffman}, {H{\"u}ntemeyer}, {Kolterman}, {Lansdell},
  {Linnemann}, {McEnery}, {Mincer}, {Moskalenko}, {Nemethy}, {Noyes}, {Porter},
  {Pretz}, {Ryan}, {Parkinson}, {Shoup}, {Sinnis}, {Smith}, {Strong},
  {Sullivan}, {Vasileiou}, {Walker}, {Williams}, \&
  {Yodh}}]{2008ApJ...688.1078A}
{Abdo}, A.~A., {Allen}, B., {Aune}, T., {et~al.} 2008, \apj, 688, 1078

\bibitem[{{Abdollahi} {et~al.}(2020){Abdollahi}, {Acero}, {Ackermann},
  {Ajello}, {Atwood}, {Axelsson}, {Baldini}, {Ballet}, {Barbiellini},
  {Bastieri}, {Becerra Gonzalez}, {Bellazzini}, {Berretta}, {Bissaldi}, {Bland
  ford}, {Bloom}, {Bonino}, {Bottacini}, {Brandt}, {Bregeon}, {Bruel},
  {Buehler}, {Burnett}, {Buson}, {Cameron}, {Caputo}, {Caraveo}, {Casandjian},
  {Castro}, {Cavazzuti}, {Charles}, {Chaty}, {Chen}, {Cheung}, {Chiaro},
  {Ciprini}, {Cohen-Tanugi}, {Cominsky}, {Coronado-Bl{\'a}zquez}, {Costantin},
  {Cuoco}, {Cutini}, {D'Ammando}, {DeKlotz}, {Torre Luque}, {de Palma},
  {Desai}, {Digel}, {Lalla}, {Mauro}, {Venere}, {Dom{\'\i}nguez}, {Dumora},
  {Dirirsa}, {Fegan}, {Ferrara}, {Franckowiak}, {Fukazawa}, {Funk}, {Fusco},
  {Gargano}, {Gasparrini}, {Giglietto}, {Giommi}, {Giordano}, {Giroletti},
  {Glanzman}, {Green}, {Grenier}, {Griffin}, {Grondin}, {Grove}, {Guiriec},
  {Harding}, {Hayashi}, {Hays}, {Hewitt}, {Horan}, {J{\'o}hannesson},
  {Johnson}, {Kamae}, {Kerr}, {Kocevski}, {Kovac'evic'}, {Kuss}, {Landriu},
  {Larsson}, {Latronico}, {Lemoine-Goumard}, {Li}, {Liodakis}, {Longo},
  {Loparco}, {Lott}, {Lovellette}, {Lubrano}, {Madejski}, {Maldera},
  {Malyshev}, {Manfreda}, {Marchesini}, {Marcotulli}, {Mart{\'\i}-Devesa},
  {Martin}, {Massaro}, {Mazziotta}, {McEnery}, {Mereu}, {Meyer}, {Michelson},
  {Mirabal}, {Mizuno}, {Monzani}, {Morselli}, {Moskalenko}, {Negro}, {Nuss},
  {Ojha}, {Omodei}, {Orienti}, {Orlando}, {Ormes}, {Palatiello}, {Paliya},
  {Paneque}, {Pei}, {Pe{\~n}a-Herazo}, {Perkins}, {Persic}, {Pesce-Rollins},
  {Petrosian}, {Petrov}, {Piron}, {Poon}, {Porter}, {Principe}, {Rain{\`o}},
  {Rando}, {Razzano}, {Razzaque}, {Reimer}, {Reimer}, {Remy}, {Reposeur},
  {Romani}, {Parkinson}, {Schinzel}, {Serini}, {Sgr{\`o}}, {Siskind}, {Smith},
  {Spandre}, {Spinelli}, {Strong}, {Suson}, {Tajima}, {Takahashi}, {Tak},
  {Thayer}, {Thompson}, {Tibaldo}, {Torres}, {Torresi}, {Valverde}, {Klaveren},
  {Zyl}, {Wood}, {Yassine}, \& {Zaharijas}}]{2020ApJS..247...33A}
{Abdollahi}, S., {Acero}, F., {Ackermann}, M., {et~al.} 2020, \apjs, 247, 33

\bibitem[{{Abeysekara} {et~al.}(2017){Abeysekara}, {Albert}, {Alfaro},
  {Alvarez}, {{\'A}lvarez}, {Arceo}, {Arteaga-Vel{\'a}zquez}, {Avila Rojas},
  {Ayala Solares}, {Barber}, {Bautista-Elivar}, {Becerril}, {Belmont-Moreno},
  {BenZvi}, {Berley}, {Bernal}, {Braun}, {Brisbois}, {Caballero-Mora},
  {Capistr{\'a}n}, {Carrami{\~n}ana}, {Casanova}, {Castillo}, {Cotti},
  {Cotzomi}, {Couti{\~n}o de Le{\'o}n}, {De Le{\'o}n}, {De la Fuente},
  {Dingus}, {DuVernois}, {D{\'{\i}}az-V{\'e}lez}, {Ellsworth}, {Engel},
  {Enr{\'{\i}}quez-Rivera}, {Fiorino}, {Fraija}, {Garc{\'{\i}}a-Gonz{\'a}lez},
  {Garfias}, {Gerhardt}, {Gonz{\'a}lez Mu{\~n}oz}, {Gonz{\'a}lez}, {Goodman},
  {Hampel-Arias}, {Harding}, {Hern{\'a}ndez}, {Hern{\'a}ndez-Almada}, {Hinton},
  {Hona}, {Hui}, {H{\"u}ntemeyer}, {Iriarte}, {Jardin-Blicq}, {Joshi},
  {Kaufmann}, {Kieda}, {Lara}, {Lauer}, {Lee}, {Lennarz}, {Vargas},
  {Linnemann}, {Longinotti}, {Luis Raya}, {Luna-Garc{\'{\i}}a},
  {L{\'o}pez-Coto}, {Malone}, {Marinelli}, {Martinez}, {Martinez-Castellanos},
  {Mart{\'{\i}}nez-Castro}, {Mart{\'{\i}}nez-Huerta}, {Matthews},
  {Miranda-Romagnoli}, {Moreno}, {Mostaf{\'a}}, {Nellen}, {Newbold}, {Nisa},
  {Noriega-Papaqui}, {Pelayo}, {Pretz}, {P{\'e}rez-P{\'e}rez}, {Ren}, {Rho},
  {Rivi{\`e}re}, {Rosa-Gonz{\'a}lez}, {Rosenberg}, {Ruiz-Velasco}, {Salazar},
  {Salesa Greus}, {Sandoval}, {Schneider}, {Schoorlemmer}, {Sinnis}, {Smith},
  {Springer}, {Surajbali}, {Taboada}, {Tibolla}, {Tollefson}, {Torres},
  {Ukwatta}, {Vianello}, {Weisgarber}, {Westerhoff}, {Wisher}, {Wood},
  {Yapici}, {Yodh}, {Younk}, {Zepeda}, {Zhou}, {Guo}, {Hahn}, {Li}, \&
  {Zhang}}]{2017Sci...358..911A}
{Abeysekara}, A.~U., {Albert}, A., {Alfaro}, R., {et~al.} 2017, Science, 358,
  911

\bibitem[{Abeysekara {et~al.}(2021)}]{HAWC:2021bvb}
Abeysekara, A.~U., {et~al.} 2021, PoS, ICRC2021, 835

\bibitem[{{Abramowski} {et~al.}(2014){Abramowski}, {Aharonian}, {Ait Benkhali},
  {Akhperjanian}, {Ang{\"u}ner}, {Backes}, {Balenderan}, {Balzer}, {Barnacka},
  {Becherini}, {Becker Tjus}, {Berge}, {Bernhard}, {Bernl{\"o}hr}, {Birsin},
  {Biteau}, {B{\"o}ttcher}, {Boisson}, {Bolmont}, {Bordas}, {Bregeon}, {Brun},
  {Brun}, {Bryan}, {Bulik}, {Carrigan}, {Casanova}, {Chadwick}, {Chakraborty},
  {Chalme-Calvet}, {Chaves}, {Chr{\'e}tien}, {Colafrancesco}, {Cologna},
  {Conrad}, {Couturier}, {Cui}, {Davids}, {Degrange}, {Deil}, {deWilt},
  {Djannati-Ata{\"\i}}, {Domainko}, {Donath}, {Drury}, {Dubus}, {Dutson},
  {Dyks}, {Dyrda}, {Edwards}, {Egberts}, {Eger}, {Espigat}, {Farnier}, {Fegan},
  {Feinstein}, {Fernandes}, {Fernandez}, {Fiasson}, {Fontaine}, {F{\"o}rster},
  {F{\"u}{\ss}ling}, {Gabici}, {Gajdus}, {Gallant}, {Garrigoux}, {Giavitto},
  {Giebels}, {Glicenstein}, {Gottschall}, {Grondin}, {Grudzi{\'n}ska},
  {Hadasch}, {H{\"a}ffner}, {Hahn}, {Harris}, {Heinzelmann}, {Henri},
  {Hermann}, {Hervet}, {Hillert}, {Hinton}, {Hofmann}, {Hofverberg}, {Holler},
  {Horns}, {Ivascenko}, {Jacholkowska}, {Jahn}, {Jamrozy}, {Janiak},
  {Jankowsky}, {Jung-Richardt}, {Kastendieck}, {Katarzy{\'n}ski}, {Katz},
  {Kaufmann}, {Kh{\'e}lifi}, {Kieffer}, {Klepser}, {Klochkov}, {Klu{\'z}niak},
  {Kolitzus}, {Komin}, {Kosack}, {Krakau}, {Krayzel}, {Kr{\"u}ger}, {Laffon},
  {Lamanna}, {Lefaucheur}, {Lefranc}, {Lemi{\`e}re}, {Lemoine-Goumard},
  {Lenain}, {Lohse}, {Lopatin}, {Lu}, {Marandon}, {Marcowith}, {Marx},
  {Maurin}, {Maxted}, {Mayer}, {McComb}, {M{\'e}hault}, {Meintjes}, {Menzler},
  {Meyer}, {Mitchell}, {Moderski}, {Mohamed}, {Mor{\^a}}, {Moulin}, {Murach},
  {de Naurois}, {Niemiec}, {Nolan}, {Oakes}, {Odaka}, {Ohm}, {Opitz},
  {Ostrowski}, {Oya}, {Panter}, {Parsons}, {Paz Arribas}, {Pekeur},
  {Pelletier}, {Petrucci}, {Peyaud}, {Pita}, {Poon}, {P{\"u}hlhofer}, {Punch},
  {Quirrenbach}, {Raab}, {Reichardt}, {Reimer}, {Reimer}, {Renaud}, {de los
  Reyes}, {Rieger}, {Romoli}, {Rosier-Lees}, {Rowell}, {Rudak}, {Rulten},
  {Sahakian}, {Salek}, {Sanchez}, {Santangelo}, {Schlickeiser},
  {Sch{\"u}ssler}, {Schulz}, {Schwanke}, {Schwarzburg}, {Schwemmer}, {Sol},
  {Spanier}, {Spengler}, {Spies}, {Stawarz}, {Steenkamp}, {Stegmann},
  {Stinzing}, {Stycz}, {Sushch}, {Tavernet}, {Tavernier}, {Taylor}, {Terrier},
  {Tluczykont}, {Trichard}, {Valerius}, {van Eldik}, {van Soelen},
  {Vasileiadis}, {Veh}, {Venter}, {Viana}, {Vincent}, {Vink}, {V{\"o}lk},
  {Volpe}, {Vorster}, {Vuillaume}, {Wagner}, {Wagner}, {Wagner}, {Ward},
  {Weidinger}, {Weitzel}, {White}, {Wierzcholska}, {Willmann}, {W{\"o}rnlein},
  {Wouters}, {Yang}, {Zabalza}, {Zaborov}, {Zacharias}, {Zdziarski}, {Zech},
  {Zechlin}, {Fukui}, \& {H.~E.~S.~S. Collaboration}}]{2014PhRvD..90l2007A}
{Abramowski}, A., {Aharonian}, F., {Ait Benkhali}, F., {et~al.} 2014, \prd, 90,
  122007

\bibitem[{{Ackermann} {et~al.}(2012){Ackermann}, {Ajello}, {Atwood}, {Baldini},
  {Ballet}, {Barbiellini}, {Bastieri}, {Bechtol}, {Bellazzini}, {Berenji},
  {Blandford}, {Bloom}, {Bonamente}, {Borgland}, {Brandt}, {Bregeon},
  {Brigida}, {Bruel}, {Buehler}, {Buson}, {Caliandro}, {Cameron}, {Caraveo},
  {Cavazzuti}, {Cecchi}, {Charles}, {Chekhtman}, {Chiang}, {Ciprini}, {Claus},
  {Cohen-Tanugi}, {Conrad}, {Cutini}, {de Angelis}, {de Palma}, {Dermer},
  {Digel}, {Silva}, {Drell}, {Drlica-Wagner}, {Falletti}, {Favuzzi}, {Fegan},
  {Ferrara}, {Focke}, {Fortin}, {Fukazawa}, {Funk}, {Fusco}, {Gaggero},
  {Gargano}, {Germani}, {Giglietto}, {Giordano}, {Giroletti}, {Glanzman},
  {Godfrey}, {Grove}, {Guiriec}, {Gustafsson}, {Hadasch}, {Hanabata},
  {Harding}, {Hayashida}, {Hays}, {Horan}, {Hou}, {Hughes}, {J{\'o}hannesson},
  {Johnson}, {Johnson}, {Kamae}, {Katagiri}, {Kataoka}, {Kn{\"o}dlseder},
  {Kuss}, {Lande}, {Latronico}, {Lee}, {Lemoine-Goumard}, {Longo}, {Loparco},
  {Lott}, {Lovellette}, {Lubrano}, {Mazziotta}, {McEnery}, {Michelson},
  {Mitthumsiri}, {Mizuno}, {Monte}, {Monzani}, {Morselli}, {Moskalenko},
  {Murgia}, {Naumann-Godo}, {Norris}, {Nuss}, {Ohsugi}, {Okumura}, {Omodei},
  {Orlando}, {Ormes}, {Paneque}, {Panetta}, {Parent}, {Pesce-Rollins},
  {Pierbattista}, {Piron}, {Pivato}, {Porter}, {Rain{\`o}}, {Rando}, {Razzano},
  {Razzaque}, {Reimer}, {Reimer}, {Sadrozinski}, {Sgr{\`o}}, {Siskind},
  {Spandre}, {Spinelli}, {Strong}, {Suson}, {Takahashi}, {Tanaka}, {Thayer},
  {Thayer}, {Thompson}, {Tibaldo}, {Tinivella}, {Torres}, {Tosti}, {Troja},
  {Usher}, {Vandenbroucke}, {Vasileiou}, {Vianello}, {Vitale}, {Waite}, {Wang},
  {Winer}, {Wood}, {Wood}, {Yang}, {Ziegler}, \&
  {Zimmer}}]{2012ApJ...750....3A}
{Ackermann}, M., {Ajello}, M., {Atwood}, W.~B., {et~al.} 2012, \apj, 750, 3

\bibitem[{{Ackermann} {et~al.}(2015){Ackermann}, {Ajello}, {Albert}, {Atwood},
  {Baldini}, {Ballet}, {Barbiellini}, {Bastieri}, {Bechtol}, {Bellazzini},
  {Bissaldi}, {Blandford}, {Bloom}, {Bottacini}, {Brandt}, {Bregeon}, {Bruel},
  {Buehler}, {Buson}, {Caliandro}, {Cameron}, {Caragiulo}, {Caraveo},
  {Cavazzuti}, {Cecchi}, {Charles}, {Chekhtman}, {Chiang}, {Chiaro}, {Ciprini},
  {Claus}, {Cohen-Tanugi}, {Conrad}, {Cuoco}, {Cutini}, {D'Ammando}, {de
  Angelis}, {de Palma}, {Dermer}, {Digel}, {Silva}, {Drell}, {Favuzzi},
  {Ferrara}, {Focke}, {Franckowiak}, {Fukazawa}, {Funk}, {Fusco}, {Gargano},
  {Gasparrini}, {Germani}, {Giglietto}, {Giommi}, {Giordano}, {Giroletti},
  {Godfrey}, {Gomez-Vargas}, {Grenier}, {Guiriec}, {Gustafsson}, {Hadasch},
  {Hayashi}, {Hays}, {Hewitt}, {Ippoliti}, {Jogler}, {J{\'o}hannesson},
  {Johnson}, {Johnson}, {Kamae}, {Kataoka}, {Kn{\"o}dlseder}, {Kuss},
  {Larsson}, {Latronico}, {Li}, {Li}, {Longo}, {Loparco}, {Lott}, {Lovellette},
  {Lubrano}, {Madejski}, {Manfreda}, {Massaro}, {Mayer}, {Mazziotta},
  {McEnery}, {Michelson}, {Mitthumsiri}, {Mizuno}, {Moiseev}, {Monzani},
  {Morselli}, {Moskalenko}, {Murgia}, {Nemmen}, {Nuss}, {Ohsugi}, {Omodei},
  {Orlando}, {Ormes}, {Paneque}, {Panetta}, {Perkins}, {Pesce-Rollins},
  {Piron}, {Pivato}, {Porter}, {Rain{\`o}}, {Rando}, {Razzano}, {Razzaque},
  {Reimer}, {Reimer}, {Reposeur}, {Ritz}, {Romani}, {S{\'a}nchez-Conde},
  {Schaal}, {Schulz}, {Sgr{\`o}}, {Siskind}, {Spandre}, {Spinelli}, {Strong},
  {Suson}, {Takahashi}, {Thayer}, {Thayer}, {Tibaldo}, {Tinivella}, {Torres},
  {Tosti}, {Troja}, {Uchiyama}, {Vianello}, {Werner}, {Winer}, {Wood}, {Wood},
  {Zaharijas}, \& {Zimmer}}]{2015ApJ...799...86A}
{Ackermann}, M., {Ajello}, M., {Albert}, A., {et~al.} 2015, \apj, 799, 86

\bibitem[{{Adriani} {et~al.}(2011){Adriani}, {Barbarino}, {Bazilevskaya},
  {Bellotti}, {Boezio}, {Bogomolov}, {Bonechi}, {Bongi}, {Bonvicini},
  {Borisov}, {Bottai}, {Bruno}, {Cafagna}, {Campana}, {Carbone}, {Carlson},
  {Casolino}, {Castellini}, {Consiglio}, {De Pascale}, {De Santis}, {De
  Simone}, {Di Felice}, {Galper}, {Gillard}, {Grishantseva}, {Jerse},
  {Karelin}, {Koldashov}, {Krutkov}, {Kvashnin}, {Leonov}, {Malakhov},
  {Malvezzi}, {Marcelli}, {Mayorov}, {Menn}, {Mikhailov}, {Mocchiutti},
  {Monaco}, {Mori}, {Nikonov}, {Osteria}, {Palma}, {Papini}, {Pearce},
  {Picozza}, {Pizzolotto}, {Ricci}, {Ricciarini}, {Rossetto}, {Sarkar},
  {Simon}, {Sparvoli}, {Spillantini}, {Stozhkov}, {Vacchi}, {Vannuccini},
  {Vasilyev}, {Voronov}, {Yurkin}, {Wu}, {Zampa}, {Zampa}, \&
  {Zverev}}]{2011Sci...332...69A}
{Adriani}, O., {Barbarino}, G.~C., {Bazilevskaya}, G.~A., {et~al.} 2011,
  Science, 332, 69

\bibitem[{{Adriani} {et~al.}(2019){Adriani}, {Akaike}, {Asano}, {Asaoka},
  {Bagliesi}, {Berti}, {Bigongiari}, {Binns}, {Bonechi}, {Bongi}, {Brogi},
  {Bruno}, {Buckley}, {Cannady}, {et~al.}}]{2019PhRvL.122r1102A}
{Adriani}, O., {Akaike}, Y., {Asano}, K., {et~al.} 2019, \prl, 122, 181102

\bibitem[{{Adriani} {et~al.}(2022){Adriani}, {Akaike}, {Asano}, {Asaoka},
  {Berti}, {Bigongiari}, {Binns}, {Bongi}, {Brogi}, {Bruno}, {Buckley},
  {Cannady}, {Castellini}, {Checchia}, {Cherry}, {Collazuol}, {Ebisawa},
  {Ficklin}, {Fuke}, {Gonzi}, {Guzik}, {Hams}, {Hibino}, {Ichimura}, {Ioka},
  {Ishizaki}, {Israel}, {Kasahara}, {Kataoka}, {Kataoka}, {Katayose}, {Kato},
  {Kawanaka}, {Kawakubo}, {Kobayashi}, {Kohri}, {Krawczynski}, {Krizmanic},
  {Maestro}, {Marrocchesi}, {Messineo}, {Mitchell}, {Miyake}, {Moiseev},
  {Mori}, {Mori}, {Motz}, {Munakata}, {Nakahira}, {Nishimura}, {de Nolfo},
  {Okuno}, {Ormes}, {Ozawa}, {Pacini}, {Papini}, {Rauch}, {Ricciarini},
  {Sakai}, {Sakamoto}, {Sasaki}, {Shimizu}, {Shiomi}, {Spillantini}, {Stolzi},
  {Sugita}, {Sulaj}, {Takita}, {Tamura}, {Terasawa}, {Torii}, {Tsunesada},
  {Uchihori}, {Vannuccini}, {Wefel}, {Yamaoka}, {Yanagita}, {Yoshida},
  {Yoshida}, {Zober}, \& {Calet Collaboration}}]{2022PhRvL.129j1102A}
---. 2022, \prl, 129, 101102

\bibitem[{{Aguilar} {et~al.}(2015{\natexlab{a}}){Aguilar}, {Aisa}, {Alpat},
  {Alvino}, {Ambrosi}, {Andeen}, {Arruda}, {Attig}, {Azzarello}, {Bachlechner},
  \& et~al.}]{2015PhRvL.115u1101A}
{Aguilar}, M., {Aisa}, D., {Alpat}, B., {et~al.} 2015{\natexlab{a}}, \prl, 115,
  211101

\bibitem[{{Aguilar} {et~al.}(2015{\natexlab{b}}){Aguilar}, {Aisa}, {Alpat},
  {Alvino}, {Ambrosi}, {Andeen}, {Arruda}, {Attig}, {Azzarello}, {Bachlechner},
  \& et~al.}]{2015PhRvL.114q1103A}
---. 2015{\natexlab{b}}, \prl, 114, 171103

\bibitem[{{Aguilar} {et~al.}(2017){Aguilar}, {Ali Cavasonza}, {Alpat},
  {Ambrosi}, {Arruda}, {Attig}, {Aupetit}, {Azzarello}, {Bachlechner}, {Barao},
  \& et~al.}]{2017PhRvL.119y1101A}
{Aguilar}, M., {Ali Cavasonza}, L., {Alpat}, B., {et~al.} 2017, \prl, 119,
  251101

\bibitem[{{Aguilar} {et~al.}(2018){Aguilar}, {Ali Cavasonza}, {Alpat},
  {Ambrosi}, {Arruda}, {Attig}, {Aupetit}, {Azzarello}, {Bachlechner}, {Barao},
  \& et~al.}]{2018PhRvL.120b1101A}
---. 2018, \prl, 120, 021101

\bibitem[{{Aguilar} {et~al.}(2019{\natexlab{a}}){Aguilar}, {Ali Cavasonza},
  {Alpat}, {Ambrosi}, {Arruda}, {Attig}, {Azzarello}, {Bachlechner}, {Barao},
  {Barrau}, {Barrin}, {Bartoloni}, {Basara}, {Ba{\c{s}}e{\v{g}}mez-du Pree},
  {Battiston}, {Becker}, {Behlmann}, {Beischer}, {Berdugo}, {Bertucci},
  {Bindi}, {de Boer}, {Bollweg}, {Borgia}, {Boschini}, {Bourquin}, {Bueno},
  {Burger}, {Burger}, {Cai}, {Capell}, {Caroff}, {Casaus}, {Castellini},
  {Cervelli}, {Chang}, {Chen}, {Chen}, {Chen}, {Cheng}, {Chou}, {Choutko},
  {Chung}, {Clark}, {Coignet}, {Consolandi}, {Contin}, {Corti}, {Crispoltoni},
  {Cui}, {Dadzie}, {Dai}, {Datta}, {Delgado}, {Della Torre}, {Demirk{\"o}z},
  {Derome}, {Di Falco}, {Di Felice}, {Dimiccoli}, {D{\'\i}az}, {von
  Doetinchem}, {Dong}, {Donnini}, {Duranti}, {Egorov}, {Eline}, {Eronen},
  {Feng}, {Fiandrini}, {Fisher}, {Formato}, {Galaktionov},
  {Garc{\'\i}a-L{\'o}pez}, {Gargiulo}, {Gast}, {Gebauer}, {Gervasi},
  {Giovacchini}, {G{\'o}mez-Coral}, {Gong}, {Goy}, {Grabski}, {Grandi},
  {Graziani}, {Guo}, {Haino}, {Han}, {He}, {Heil}, {Hsieh}, {Huang}, {Huang},
  {Incagli}, {Jia}, {Jinchi}, {Kanishev}, {Khiali}, {Kirn}, {Konak}, {Kounina},
  {Kounine}, {Koutsenko}, {Kulemzin}, {La Vacca}, {Laudi}, {Laurenti},
  {Lazzizzera}, {Lebedev}, {Lee}, {Lee}, {Leluc}, {Li}, {Li}, {Li}, {Li},
  {Light}, {Lin}, {Lippert}, {Liu}, {Liu}, {Liu}, {Lu}, {Lu}, {Luebelsmeyer},
  {Luo}, {Luo}, {Luo}, {Lyu}, {Machate}, {Ma{\~n}{\'a}}, {Mar{\'\i}n},
  {Martin}, {Mart{\'\i}nez}, {Masi}, {Maurin}, {Menchaca-Rocha}, {Meng}, {Mo},
  {Molero}, {Mott}, {Mussolin}, {Nelson}, {Ni}, {Nikonov}, {Nozzoli}, {Oliva},
  {Orcinha}, {Palermo}, {Palmonari}, {Paniccia}, {Pashnin}, {Pauluzzi},
  {Pensotti}, {Perrina}, {Phan}, {Picot-Clemente}, {Plyaskin}, {Pohl},
  {Poireau}, {Popkow}, {Quadrani}, {Qi}, {Qin}, {Qu}, {Rancoita}, {Rapin},
  {Conde}, {Rosier-Lees}, {Rozhkov}, {Rozza}, {Sagdeev}, {Solano}, {Schael},
  {Schmidt}, {von Dratzig}, {Schwering}, {Seo}, {Shan}, {Shi}, {Siedenburg},
  {Song}, {Sun}, {Tacconi}, {Tang}, {Tang}, {Tian}, {Ting}, {Ting},
  {Tomassetti}, {Torsti}, {Urban}, {Vagelli}, {Valente}, {Valtonen}, {Acosta},
  {Vecchi}, {Velasco}, {Vialle}, {Viz{\'a}n}, {Wang}, {Wang}, {Wang}, {Wang},
  {Wang}, {Wang}, {Wei}, {Weng}, {Wu}, {Xiong}, {Xu}, {Yan}, {Yang}, {Yi},
  {Yu}, {Yu}, {Zannoni}, {Zeissler}, {Zhang}, {Zhang}, {Zhang}, {Zhang},
  {Zhao}, {Zheng}, {Zhuang}, {Zhukov}, {Zichichi}, {Zimmermann}, {Zuccon}, \&
  {AMS Collaboration}}]{2019PhRvL.122j1101A}
---. 2019{\natexlab{a}}, \prl, 122, 101101

\bibitem[{{Aguilar} {et~al.}(2019{\natexlab{b}}){Aguilar}, {Ali Cavasonza},
  {Ambrosi}, {Arruda}, {Attig}, {Azzarello}, {Bachlechner}, {Barao}, {Barrau},
  {Barrin}, {Bartoloni}, {Basara}, {Ba{\c{s}}e{\v{g}}mez-du Pree}, {Battiston},
  {Becker}, {Behlmann}, {Beischer}, {Berdugo}, {Bertucci}, {Bindi}, {de Boer},
  {Bollweg}, {Borgia}, {Boschini}, {Bourquin}, {Bueno}, {Burger}, {Burger},
  {Cai}, {Capell}, {Caroff}, {Casaus}, {Castellini}, {Cervelli}, {Chang},
  {Chen}, {Chen}, {Chen}, {Cheng}, {Chou}, {Choutko}, {Chung}, {Clark},
  {Coignet}, {Consolandi}, {Contin}, {Corti}, {Crispoltoni}, {Cui}, {Dadzie},
  {Dai}, {Datta}, {Delgado}, {Della Torre}, {Demirk{\"o}z}, {Derome}, {Di
  Falco}, {Dimiccoli}, {D{\'\i}az}, {von Doetinchem}, {Dong}, {Donnini},
  {Duranti}, {Egorov}, {Eline}, {Eronen}, {Feng}, {Fiandrini}, {Fisher},
  {Formato}, {Galaktionov}, {Garc{\'\i}a-L{\'o}pez}, {Gargiulo}, {Gast},
  {Gebauer}, {Gervasi}, {Giovacchini}, {G{\'o}mez-Coral}, {Gong}, {Goy},
  {Grabski}, {Grandi}, {Graziani}, {Guo}, {Haino}, {Han}, {He}, {Heil},
  {Hsieh}, {Huang}, {Huang}, {Incagli}, {Jia}, {Jinchi}, {Kanishev}, {Khiali},
  {Kirn}, {Konak}, {Kounina}, {Kounine}, {Koutsenko}, {Kulemzin}, {La Vacca},
  {Laudi}, {Laurenti}, {Lazzizzera}, {Lebedev}, {Lee}, {Lee}, {Leluc}, {Li},
  {Li}, {Li}, {Li}, {Light}, {Lin}, {Lippert}, {Liu}, {Liu}, {Liu}, {Lu}, {Lu},
  {Luebelsmeyer}, {Luo}, {Luo}, {Luo}, {Lyu}, {Machate}, {Ma{\~n}{\'a}},
  {Mar{\'\i}n}, {Martin}, {Mart{\'\i}nez}, {Masi}, {Maurin}, {Menchaca-Rocha},
  {Meng}, {Mo}, {Molero}, {Mott}, {Mussolin}, {Nelson}, {Ni}, {Nikonov},
  {Nozzoli}, {Oliva}, {Orcinha}, {Palermo}, {Palmonari}, {Paniccia}, {Pashnin},
  {Pauluzzi}, {Pensotti}, {Perrina}, {Phan}, {Picot-Clemente}, {Plyaskin},
  {Pohl}, {Poireau}, {Popkow}, {Quadrani}, {Qi}, {Qin}, {Qu}, {Rancoita},
  {Rapin}, {Conde}, {Rosier-Lees}, {Rozhkov}, {Rozza}, {Sagdeev}, {Solano},
  {Schael}, {Schmidt}, {Schulz von Dratzig}, {Schwering}, {Seo}, {Shan}, {Shi},
  {Siedenburg}, {Song}, {Sun}, {Tacconi}, {Tang}, {Tang}, {Tian}, {Ting},
  {Ting}, {Tomassetti}, {Torsti}, {Urban}, {Vagelli}, {Valente}, {Valtonen},
  {V{\'a}zquez Acosta}, {Vecchi}, {Velasco}, {Vialle}, {Viz{\'a}n}, {Wang},
  {Wang}, {Wang}, {Wang}, {Wang}, {Wang}, {Wei}, {Weng}, {Wu}, {Xiong}, {Xu},
  {Yan}, {Yang}, {Yi}, {Yu}, {Yu}, {Zannoni}, {Zeissler}, {Zhang}, {Zhang},
  {Zhang}, {Zhang}, {Zhao}, {Zheng}, {Zhuang}, {Zhukov}, {Zichichi},
  {Zimmermann}, {Zuccon}, \& {AMS Collaboration}}]{2019PhRvL.122d1102A}
{Aguilar}, M., {Ali Cavasonza}, L., {Ambrosi}, G., {et~al.} 2019{\natexlab{b}},
  \prl, 122, 041102

\bibitem[{{Aguilar} {et~al.}(2021){Aguilar}, {Ali Cavasonza}, {Ambrosi},
  {Arruda}, {Attig}, {Barao}, {Barrin}, {Bartoloni}, {Ba{\c{s}}e{\u{g}}mez-du
  Pree}, {Bates}, {Battiston}, {Behlmann}, {Beischer}, {Berdugo}, {Bertucci},
  {Bindi}, {de Boer}, {Bollweg}, {Borgia}, {Boschini}, {Bourquin}, {Bueno},
  {Burger}, {Burger}, {Burmeister}, {Cai}, {Capell}, {Casaus}, {Castellini},
  {Cervelli}, {Chang}, {Chen}, {Chen}, {Chen}, {Cheng}, {Chou}, {Chouridou},
  {Choutko}, {Chung}, {Clark}, {Coignet}, {Consolandi}, {Contin}, {Corti},
  {Cui}, {Dadzie}, {Dai}, {Delgado}, {Della Torre}, {Demirk{\"o}z}, {Derome},
  {Di Falco}, {Di Felice}, {D{\'\i}az}, {Dimiccoli}, {von Doetinchem}, {Dong},
  {Donnini}, {Duranti}, {Egorov}, {Eline}, {Feng}, {Fiandrini}, {Fisher},
  {Formato}, {Freeman}, {Galaktionov}, {G{\'a}mez}, {Garc{\'\i}a-L{\'o}pez},
  {Gargiulo}, {Gast}, {Gebauer}, {Gervasi}, {Giovacchini}, {G{\'o}mez-Coral},
  {Gong}, {Goy}, {Grabski}, {Grandi}, {Graziani}, {Guo}, {Haino}, {Han},
  {Hashmani}, {He}, {Heber}, {Hsieh}, {Hu}, {Huang}, {Hungerford}, {Incagli},
  {Jang}, {Jia}, {Jinchi}, {Kanishev}, {Khiali}, {Kim}, {Kirn}, {Konyushikhin},
  {Kounina}, {Kounine}, {Koutsenko}, {Kuhlman}, {Kulemzin}, {La Vacca},
  {Laudi}, {Laurenti}, {Lazzizzera}, {Lebedev}, {Lee}, {Lee}, {Leluc}, {Li},
  {Li}, {Li}, {Li}, {Li}, {Li}, {Light}, {Lin}, {Lippert}, {Liu}, {Lu}, {Lu},
  {Luebelsmeyer}, {Luo}, {Lyu}, {Machate}, {Ma{\~n}{\'a}}, {Mar{\'\i}n},
  {Marquardt}, {Martin}, {Mart{\'\i}nez}, {Masi}, {Maurin}, {Menchaca-Rocha},
  {Meng}, {Mo}, {Molero}, {Mott}, {Mussolin}, {Ni}, {Nikonov}, {Nozzoli},
  {Oliva}, {Orcinha}, {Palermo}, {Palmonari}, {Paniccia}, {Pashnin},
  {Pauluzzi}, {Pensotti}, {Phan}, {Plyaskin}, {Pohl}, {Porter}, {Qi}, {Qin},
  {Qu}, {Quadrani}, {Rancoita}, {Rapin}, {Reina Conde}, {Rosier-Lees},
  {Rozhkov}, {Rozza}, {Sagdeev}, {Schael}, {Schmidt}, {Schulz von Dratzig},
  {Schwering}, {Seo}, {Shan}, {Shi}, {Siedenburg}, {Solano}, {Song},
  {Sonnabend}, {Sun}, {Sun}, {Tacconi}, {Tang}, {Tang}, {Tian}, {Ting}, {Ting},
  {Tomassetti}, {Torsti}, {T{\"u}ys{\"u}z}, {Urban}, {Usoskin}, {Vagelli},
  {Vainio}, {Valente}, {Valtonen}, {V{\'a}zquez Acosta}, {Vecchi}, {Velasco},
  {Vialle}, {Wang}, {Wang}, {Wang}, {Wang}, {Wang}, {Wang}, {Wei}, {Weng},
  {Wu}, {Xiong}, {Xu}, {Yan}, {Yang}, {Yi}, {Yu}, {Yu}, {Zannoni}, {Zhang},
  {Zhang}, {Zhang}, {Zhang}, {Zhang}, {Zhao}, {Zheng}, {Zhuang}, {Zhukov},
  {Zichichi}, {Zimmermann}, {Zuccon}, \& {AMS
  Collaboration}}]{2021PhR...894....1A}
---. 2021, \physrep, 894, 1

\bibitem[{{Aharonian} {et~al.}(2021){Aharonian}, {An}, {Axikegu}, {Bai}, {Bao},
  {Bastieri}, {Bi}, {Bi}, {Cai}, {Cai}, {Cao}, {Cao}, {Chang}, {Chang},
  {Chang}, {Chen}, {Chen}, {Chen}, {Chen}, {Chen}, {Chen}, {Chen}, {Chen},
  {Chen}, {Chen}, {Chen}, {Chen}, {Chen}, {Cheng}, {Cheng}, {Cui}, {Cui},
  {Cui}, {Dai}, {Dai}, {Dai}, {Danzengluobu}, {Della Volpe}, {D'Ettorre
  Piazzoli}, {Dong}, {Fan}, {Fan}, {Fan}, {Fang}, {Fang}, {Feng}, {Feng},
  {Feng}, {Feng}, {Gao}, {Gao}, {Gao}, {Gao}, {Ge}, {Geng}, {Gong}, {Gou},
  {Gu}, {Guo}, {Guo}, {Guo}, {Guo}, {Han}, {He}, {He}, {He}, {He}, {He}, {He},
  {Heller}, {Hor}, {Hou}, {Hou}, {Hu}, {Hu}, {Hu}, {Hu}, {Huang}, {Huang},
  {Huang}, {Huang}, {Huang}, {Ji}, {Ji}, {Jia}, {Jiang}, {Jiang}, {Jin},
  {Kuleshov}, {Levochkin}, {Li}, {Li}, {Li}, {Li}, {Li}, {Li}, {Li}, {Li},
  {Li}, {Li}, {Li}, {Li}, {Li}, {Li}, {Li}, {Li}, {Li}, {Liang}, {Liang},
  {Lin}, {Liu}, {Liu}, {Liu}, {Liu}, {Liu}, {Liu}, {Liu}, {Liu}, {Liu}, {Liu},
  {Liu}, {Liu}, {Liu}, {Liu}, {Liu}, {Long}, {Lu}, {Lv}, {Ma}, {Ma}, {Ma},
  {Mao}, {Masood}, {Mitthumsiri}, {Montaruli}, {Nan}, {Pang},
  {Pattarakijwanich}, {Pei}, {Qi}, {Ruffolo}, {Rulev}, {S{\'a}iz}, {Shao},
  {Shchegolev}, {Sheng}, {Shi}, {Song}, {Stenkin}, {Stepanov}, {Sun}, {Sun},
  {Sun}, {Tam}, {Tang}, {Tian}, {Wang}, {Wang}, {Wang}, {Wang}, {Wang}, {Wang},
  {Wang}, {Wang}, {Wang}, {Wang}, {Wang}, {Wang}, {Wang}, {Wang}, {Wang},
  {Wang}, {Wang}, {Wang}, {Wang}, {Wang}, {Wang}, {Wei}, {Wei}, {Wei}, {Wen},
  {Wu}, {Wu}, {Wu}, {Wu}, {Wu}, {Xi}, {Xia}, {Xia}, {Xiang}, {Xiao}, {Xiao},
  {Xin}, {Xin}, {Xing}, {Xu}, {Xu}, {Xue}, {Yan}, {Yang}, {Yang}, {Yang},
  {Yang}, {Yang}, {Yang}, {Yang}, {Yao}, {Yao}, {Ye}, {Yin}, {Yin}, {You},
  {You}, {Yu}, {Yuan}, {Zeng}, {Zeng}, {Zeng}, {Zeng}, {Zha}, {Zhai}, {Zhang},
  {Zhang}, {Zhang}, {Zhang}, {Zhang}, {Zhang}, {Zhang}, {Zhang}, {Zhang},
  {Zhang}, {Zhang}, {Zhang}, {Zhang}, {Zhang}, {Zhang}, {Zhang}, {Zhang},
  {Zhang}, {Zhang}, {Zhao}, {Zhao}, {Zhao}, {Zhao}, {Zhao}, {Zheng}, {Zheng},
  {Zhou}, {Zhou}, {Zhou}, {Zhou}, {Zhou}, {Zhou}, {Zhu}, {Zhu}, {Zhu}, {Zhu},
  {Zuo}, {LHAASO Collaboration}, \& {Huang}}]{2021PhRvL.126x1103A}
{Aharonian}, F., {An}, Q., {Axikegu}, Bai, L.~X., {et~al.} 2021, \prl, 126,
  241103

\bibitem[{{Aharonian} \& {Atoyan}(2000)}]{2000A&A...362..937A}
{Aharonian}, F.~A., \& {Atoyan}, A.~M. 2000, \aap, 362, 937

\bibitem[{{Ahn} {et~al.}(2010){Ahn}, {Allison}, {Bagliesi}, {Beatty},
  {Bigongiari}, {Childers}, {Conklin}, {Coutu}, {DuVernois}, {Ganel}, {Han},
  {Jeon}, {Kim}, {Lee}, {Lutz}, {Maestro}, {Malinin}, {Marrocchesi}, {Minnick},
  {Mognet}, {Nam}, {Nam}, {Nutter}, {Park}, {Park}, {Seo}, {Sina}, {Wu},
  {Yang}, {Yoon}, {Zei}, \& {Zinn}}]{2010ApJ...714L..89A}
{Ahn}, H.~S., {Allison}, P., {Bagliesi}, M.~G., {et~al.} 2010, \apjl, 714, L89

\bibitem[{{Alemanno} {et~al.}(2021){Alemanno}, {An}, {Azzarello}, {Barbato},
  {Bernardini}, {Bi}, {Cai}, {Catanzani}, {Chang}, {Chen}, {Chen}, {Chen},
  {Cui}, {Cui}, {Cui}, {Dai}, {D'Amone}, {de Benedittis}, {de Mitri}, {de
  Palma}, {Deliyergiyev}, {di Santo}, {Dong}, {Dong}, {Donvito}, {Droz},
  {Duan}, {Duan}, {D'Urso}, {Fan}, {Fan}, {Fang}, {Fang}, {Feng}, {Feng},
  {Fusco}, {Gao}, {Gargano}, {Gong}, {Gong}, {Guo}, {Guo}, {Guo}, {Han}, {Hu},
  {Huang}, {Huang}, {Huang}, {Ionica}, {Jiang}, {Kong}, {Kotenko}, {Kyratzis},
  {Lei}, {Li}, {Li}, {Li}, {Li}, {Liang}, {Liu}, {Liu}, {Liu}, {Liu}, {Liu},
  {Liu}, {Loparco}, {Luo}, {Ma}, {Ma}, {Ma}, {Ma}, {Marsella}, {Mazziotta},
  {Mo}, {Niu}, {Pan}, {Parenti}, {Peng}, {Peng}, {Perrina}, {Qiao}, {Rao},
  {Ruina}, {Salinas}, {Shang}, {Shen}, {Shen}, {Shen}, {Silveri}, {Song},
  {Stolpovskiy}, {Su}, {Su}, {Sun}, {Surdo}, {Teng}, {Tykhonov}, {Wang},
  {Wang}, {Wang}, {Wang}, {Wang}, {Wang}, {Wang}, {Wang}, {Wang}, {Wei}, {Wei},
  {Wei}, {Wen}, {Wu}, {Wu}, {Wu}, {Wu}, {Wu}, {Xia}, {Xu}, {Xu}, {Xu}, {Xu},
  {Xue}, {Yang}, {Yang}, {Yang}, {Yao}, {Yu}, {Yuan}, {Yuan}, {Yue}, {Zang},
  {Zhang}, {Zhang}, {Zhang}, {Zhang}, {Zhang}, {Zhang}, {Zhang}, {Zhang},
  {Zhang}, {Zhang}, {Zhao}, {Zhao}, {Zhao}, {Zhou}, {Zhu}, \& {Dampe
  Collaboration}}]{2021PhRvL.126t1102A}
{Alemanno}, F., {An}, Q., {Azzarello}, P., {et~al.} 2021, \prl, 126, 201102

\bibitem[{{Alemanno} {et~al.}(2022){Alemanno}, {An}, {Azzarello}, {Carla
  Tiziana Barbato}, {Bernardini}, {Bi}, {Cai}, {Casilli}, {Catanzani}, {Chang},
  {et~al.}}]{2022SciBu..67.2162D}
---. 2022, Science Bulletin, 67, 2162

\bibitem[{{Amenomori} {et~al.}(2021){Amenomori}, {Bao}, {Bi}, {Chen}, {Chen},
  {Chen}, {Chen}, {Chen}, {Cirennima}, {Cui}, {Danzengluobu}, {Ding}, {Fang},
  {Fang}, {Feng}, {Feng}, {Feng}, {Gao}, {Gou}, {Guo}, {He}, {He}, {Hibino},
  {Hotta}, {Hu}, {Hu}, {Huang}, {Jia}, {Jiang}, {Jin}, {Kajino}, {Kasahara},
  {Katayose}, {Kato}, {Kato}, {Kawata}, {Kozai}, {Labaciren}, {Le}, {Li}, {Li},
  {Li}, {Lin}, {Liu}, {Liu}, {Liu}, {Liu}, {Lou}, {Lu}, {Meng}, {Mitsui},
  {Munakata}, {Nakamura}, {Nanjo}, {Nishizawa}, {Ohnishi}, {Ohta}, {Ozawa},
  {Qian}, {Qu}, {Saito}, {Sakata}, {Sako}, {Sengoku}, {Shao}, {Shibata},
  {Shiomi}, {Sugimoto}, {Takita}, {Tan}, {Tateyama}, {Torii}, {Tsuchiya},
  {Udo}, {Wang}, {Wu}, {Xue}, {Yagisawa}, {Yamamoto}, {Yang}, {Yuan}, {Zhai},
  {Zhang}, {Zhang}, {Zhang}, {Zhang}, {Zhang}, {Zhang}, {Zhang},
  {Zhaxisangzhu}, {Zhou}, \& {Tibet AS {\ensuremath{\gamma}}
  Collaboration}}]{2021PhRvL.126n1101A}
{Amenomori}, M., {Bao}, Y.~W., {Bi}, X.~J., {et~al.} 2021, \prl, 126, 141101

\bibitem[{{An} {et~al.}(2019){An}, {Asfandiyarov}, {Azzarello}, {Bernardini},
  {Bi}, {Cai}, {Chang}, {Chen}, {Chen}, {Chen}, {Chen}, {Cui}, {Cui}, {Dai},
  {D'Amone}, {De Benedittis}, {De Mitri}, {Di Santo}, {Ding}, {Dong}, {Dong},
  {Dong}, {Donvito}, {Droz}, {Duan}, {Duan}, {D'Urso}, {Fan}, {Fan}, {Fang},
  {Feng}, {Feng}, {Fusco}, {Gallo}, {Gan}, {Gao}, {Gargano}, {Gong}, {Gong},
  {Guo}, {Guo}, {Guo}, {Han}, {Hu}, {Huang}, {Huang}, {Huang}, {Ionica},
  {Jiang}, {Jin}, {Kong}, {Lei}, {Li}, {Li}, {Li}, {Li}, {Li}, {Liang},
  {Liang}, {Liao}, {Liu}, {Liu}, {Liu}, {Liu}, {Liu}, {Liu}, {Loparco}, {Luo},
  {Ma}, {Ma}, {Ma}, {Ma}, {Ma}, {Marsella}, {Mazziotta}, {Mo}, {Niu}, {Pan},
  {Peng}, {Peng}, {Qiao}, {Rao}, {Salinas}, {Shang}, {Shen}, {Shen}, {Shen},
  {Song}, {Su}, {Su}, {Sun}, {Surdo}, {Teng}, {Tykhonov}, {Vitillo}, {Wang},
  {Wang}, {Wang}, {Wang}, {Wang}, {Wang}, {Wang}, {Wang}, {Wang}, {Wang},
  {Wang}, {Wang}, {Wang}, {Wei}, {Wei}, {Wei}, {Wen}, {Wu}, {Wu}, {Wu}, {Wu},
  {Wu}, {Xi}, {Xia}, {Xu}, {Xu}, {Xu}, {Xu}, {Xue}, {Yang}, {Yang}, {Yang},
  {Yang}, {Yao}, {Yu}, {Yuan}, {Yue}, {Zang}, {Zhang}, {Zhang}, {Zhang},
  {Zhang}, {Zhang}, {Zhang}, {Zhang}, {Zhang}, {Zhang}, {Zhang}, {Zhang},
  {Zhang}, {Zhang}, {Zhao}, {Zhao}, {Zhao}, {Zhou}, {Zhou}, {Zhu}, {Zhu}, \&
  {Zimmer}}]{2019SciA....5.3793A}
{An}, Q., {Asfandiyarov}, R., {Azzarello}, P., {et~al.} 2019, Science Advances,
  5, eaax3793

\bibitem[{{Antoni} {et~al.}(2005){Antoni}, {Apel}, {Badea}, {Bekk}, {Bercuci},
  {Bl{\"u}mer}, {Bozdog}, {Brancus}, {Chilingarian}, {Daumiller}, {Doll},
  {Engel}, {Engler}, {Fe{\ss}ler}, {Gils}, {Glasstetter}, {Haungs}, {Heck},
  {H{\"o}randel}, {Kampert}, {Klages}, {Maier}, {Mathes}, {Mayer}, {Milke},
  {M{\"u}ller}, {Obenland}, {Oehlschl{\"a}ger}, {Ostapchenko}, {Petcu},
  {Rebel}, {Risse}, {Risse}, {Roth}, {Schatz}, {Schieler}, {Scholz}, {Thouw},
  {Ulrich}, {van Buren}, {Vardanyan}, {Weindl}, {Wochele}, \&
  {Zabierowski}}]{2005APh....24....1A}
{Antoni}, T., {Apel}, W.~D., {Badea}, A.~F., {et~al.} 2005, Astropart. Phys.,
  24, 1

\bibitem[{{Apel} {et~al.}(2013){Apel}, {Arteaga-Vel{\'a}zquez}, {Bekk},
  {Bertaina}, {Bl{\"u}mer}, {Bozdog}, {Brancus}, {Cantoni}, {Chiavassa},
  {Cossavella}, {Daumiller}, {de Souza}, {Di Pierro}, {Doll}, {Engel},
  {Engler}, {Finger}, {Fuchs}, {Fuhrmann}, {Gils}, {Glasstetter}, {Grupen},
  {Haungs}, {Heck}, {H{\"o}randel}, {Huber}, {Huege}, {Kampert}, {Kang},
  {Klages}, {Link}, {{\L}uczak}, {Ludwig}, {Mathes}, {Mayer}, {Melissas},
  {Milke}, {Mitrica}, {Morello}, {Oehlschl{\"a}ger}, {Ostapchenko}, {Palmieri},
  {Petcu}, {Pierog}, {Rebel}, {Roth}, {Schieler}, {Schoo}, {Schr{\"o}der},
  {Sima}, {Toma}, {Trinchero}, {Ulrich}, {Weindl}, {Wochele}, {Wommer}, \&
  {Zabierowski}}]{2013APh....47...54A}
{Apel}, W.~D., {Arteaga-Vel{\'a}zquez}, J.~C., {Bekk}, K., {et~al.} 2013,
  Astroparticle Physics, 47, 54

\bibitem[{{Atkin} {et~al.}(2018){Atkin}, {Bulatov}, {Dorokhov}, {Gorbunov},
  {Filippov}, {Grebenyuk}, {Karmanov}, {Kovalev}, {Kudryashov}, {Kurganov},
  {Merkin}, {Panov}, {Podorozhny}, {Polkov}, {Porokhovoy}, {Shumikhin},
  {Tkachenko}, {Tkachev}, {Turundaevskiy}, {Vasiliev}, \&
  {Voronin}}]{2018JETPL.108....5A}
{Atkin}, E., {Bulatov}, V., {Dorokhov}, V., {et~al.} 2018, Soviet Journal of
  Experimental and Theoretical Physics Letters, 108, 5

\bibitem[{{Bartoli} {et~al.}(2015){Bartoli}, {Bernardini}, {Bi}, {Branchini},
  {Budano}, {Camarri}, {Cao}, {Cardarelli}, {Catalanotti}, {Chen}, {Chen},
  {Creti}, {Cui}, {Dai}, {D'Amone}, {Danzengluobu}, {De Mitri}, {D'Ettorre
  Piazzoli}, {Di Girolamo}, {Di Sciascio}, {Feng}, {Feng}, {Feng}, {Gou},
  {Guo}, {He}, {Hu}, {Hu}, {Iacovacci}, {Iuppa}, {Jia}, {Labaciren}, {Li},
  {Liguori}, {Liu}, {Liu}, {Liu}, {Lu}, {Ma}, {Ma}, {Mancarella}, {Mari},
  {Marsella}, {Martello}, {Mastroianni}, {Montini}, {Ning}, {Panareo},
  {Perrone}, {Pistilli}, {Ruggieri}, {Salvini}, {Santonico}, {Shen}, {Sheng},
  {Shi}, {Surdo}, {Tan}, {Vallania}, {Vernetto}, {Vigorito}, {Wang}, {Wu},
  {Wu}, {Xue}, {Yang}, {Yang}, {Yao}, {Yuan}, {Zha}, {Zhang}, {Zhang}, {Zhang},
  {Zhang}, {Zhao}, {Zhaxiciren}, {Zhaxisangzhu}, {Zhou}, {Zhu}, {Zhu}, {Zizzi},
  \& {ARGO-YBJ Collaboration}}]{2015ApJ...806...20B}
{Bartoli}, B., {Bernardini}, P., {Bi}, X.~J., {et~al.} 2015, \apj, 806, 20

\bibitem[{{Breuhaus} {et~al.}(2022){Breuhaus}, {Hinton}, {Joshi}, {Reville}, \&
  {Schoorlemmer}}]{2022A&A...661A..72B}
{Breuhaus}, M., {Hinton}, J.~A., {Joshi}, V., {Reville}, B., \& {Schoorlemmer},
  H. 2022, \aap, 661, A72

\bibitem[{{Bronfman} {et~al.}(1988){Bronfman}, {Cohen}, {Alvarez}, {May}, \&
  {Thaddeus}}]{1988ApJ...324..248B}
{Bronfman}, L., {Cohen}, R.~S., {Alvarez}, H., {May}, J., \& {Thaddeus}, P.
  1988, \apj, 324, 248

\bibitem[{{Cao} {et~al.}(2023{\natexlab{a}}){Cao}, {Aharonian}, {An},
  {Axikegu}, {Bai}, {Bao}, {Bastieri}, {Bi}, {Bi}, {Cai}, {Cao}, {Cao}, {Cao},
  {Chang}, {Chang}, {et~al.}}]{LHAASO-diffuse-2023}
{Cao}, Z., {Aharonian}, F., {An}, Q., {et~al.} 2023{\natexlab{a}}, arXiv
  e-prints, arXiv:2305.05372

\bibitem[{{Cao} {et~al.}(2023{\natexlab{b}}){Cao}, {Aharonian}, {An},
  {Axikegu}, {Bai}, {Bao}, {Bastieri}, {Bi}, {Bi}, {Cai}, \&
  et~al.}]{2023arXiv230517030C}
---. 2023{\natexlab{b}}, arXiv e-prints, arXiv:2305.17030

\bibitem[{{Choi} {et~al.}(2022){Choi}, {Seo}, {Aggarwal}, {Amare},
  {Angelaszek}, {Bowman}, {Chen}, {Copley}, {Derome}, {Eraud}, {Falana},
  {Gerrety}, {Han}, {Huh}, {Haque}, {Hwang}, {Hyun}, {Jeon}, {Jeon}, {Jeong},
  {Kang}, {Kim}, {Kim}, {Kim}, {Lee}, {Lee}, {Lee}, {Lu}, {Lundquist}, {Lutz},
  {Menchaca-Rocha}, {Ofoha}, {Park}, {Park}, {Park}, {Picot-Clemente},
  {Scrandis}, {Smith}, {Takeishi}, {Vedenkin}, {Walpole}, {Weinmann}, {Wu},
  {Wu}, {Yin}, {Yoon}, \& {Zhang}}]{2022ApJ...940..107C}
{Choi}, G.~H., {Seo}, E.~S., {Aggarwal}, S., {et~al.} 2022, \apj, 940, 107

\bibitem[{{Cordes} {et~al.}(1991){Cordes}, {Weisberg}, {Frail}, {Spangler}, \&
  {Ryan}}]{1991Natur.354..121C}
{Cordes}, J.~M., {Weisberg}, J.~M., {Frail}, D.~A., {Spangler}, S.~R., \&
  {Ryan}, M. 1991, \nat, 354, 121

\bibitem[{{Cummings} {et~al.}(2016){Cummings}, {Stone}, {Heikkila}, {Lal},
  {Webber}, {J{\'o}hannesson}, {Moskalenko}, {Orlando}, \&
  {Porter}}]{2016ApJ...831...18C}
{Cummings}, A.~C., {Stone}, E.~C., {Heikkila}, B.~C., {et~al.} 2016, \apj, 831,
  18

\bibitem[{{DAMPE Collaboration} {et~al.}(2017){DAMPE Collaboration}, {Ambrosi},
  {An}, {Asfandiyarov}, {Azzarello}, {Bernardini}, {Bertucci}, {Cai}, {Chang},
  {Chen}, {Chen}, {Chen}, {Chen}, {Cui}, {Cui}, {D'Amone}, {de Benedittis}, {De
  Mitri}, {di Santo}, {Dong}, {Dong}, {Dong}, {Dong}, {Donvito}, {Droz},
  {Duan}, {Duan}, {Duranti}, {D'Urso}, {Fan}, {Fan}, {Fang}, {Feng}, {Feng},
  {Fusco}, {Gallo}, {Gan}, {Gao}, {Gao}, {Gargano}, {Garrappa}, {Gong}, {Gong},
  {Guo}, {Guo}, {Hu}, {Huang}, {Huang}, {Ionica}, {Jiang}, {Jiang}, {Jin},
  {Kong}, {Lei}, {Li}, {Li}, {Li}, {Li}, {Liang}, {Liang}, {Liao}, {Liu},
  {Liu}, {Liu}, {Liu}, {Liu}, {Loparco}, {Ma}, {Ma}, {Ma}, {Ma}, {Ma}, {Ma},
  {Marsella}, {Mazziotta}, {Mo}, {Niu}, {Peng}, {Peng}, {Qiao}, {Rao},
  {Salinas}, {Shang}, {H.~Shen}, {Shen}, {Shen}, {Song}, {Su}, {Su}, {Sun},
  {Surdo}, {Teng}, {Tian}, {Tykhonov}, {Vagelli}, {Vitillo}, {Wang}, {Wang},
  {Wang}, {Wang}, {Wang}, {Wang}, {Wang}, {Wang}, {Wang}, {Wang}, {Wang},
  {Wang}, {Wen}, {Wang}, {Wei}, {Wei}, {Wei}, {Wu}, {Wu}, {Wu}, {Wu}, {Wu},
  {Xi}, {Xia}, {Xin}, {Xu}, {Xu}, {Xu}, {Xue}, {Yang}, {Yang}, {Yang}, {Yang},
  {Yao}, {Yu}, {Yuan}, {Yue}, {Zang}, {Zhang}, {Zhang}, {Zhang}, {Zhang},
  {Zhang}, {Zhang}, {Zhang}, {Zhang}, {Zhang}, {Zhang}, {Zhang}, {Zhang},
  {Zhang}, {Zhang}, {Zhang}, {Zhang}, {Zhang}, {Zhao}, {Zhao}, {Zhao}, {Zhou},
  {Zhou}, {Zhu}, {Zhu}, \& {Zimmer}}]{2017Natur.552...63D}
{DAMPE Collaboration}, {Ambrosi}, G., {An}, Q., {et~al.} 2017, \nat, 552, 63

\bibitem[{{De La Torre Luque} {et~al.}(2023){De La Torre Luque}, {Gaggero},
  {Grasso}, {Fornieri}, {Egberts}, {Steppa}, \& {Evoli}}]{2023A&A...672A..58D}
{De La Torre Luque}, P., {Gaggero}, D., {Grasso}, D., {et~al.} 2023, \aap, 672,
  A58

\bibitem[{{Dzhatdoev}(2021)}]{2021arXiv210402838D}
{Dzhatdoev}, T. 2021, arXiv e-prints, arXiv:2104.02838

\bibitem[{Engel {et~al.}(2011)Engel, Heck, \& Pierog}]{Engel:2011zzb}
Engel, R., Heck, D., \& Pierog, T. 2011, Ann. Rev. Nucl. Part. Sci., 61, 467

\bibitem[{{Fang} \& {Murase}(2021)}]{2021ApJ...919...93F}
{Fang}, K., \& {Murase}, K. 2021, \apj, 919, 93

\bibitem[{{Fang} \& {Murase}(2023)}]{2023arXiv230702905F}
---. 2023, arXiv e-prints, arXiv:2307.02905

\bibitem[{{G{\'e}nolini} {et~al.}(2017){G{\'e}nolini}, {Serpico}, {Boudaud},
  {Caroff}, {Poulin}, {Derome}, {Lavalle}, {Maurin}, {Poireau}, {Rosier},
  {Salati}, \& {Vecchi}}]{2017PhRvL.119x1101G}
{G{\'e}nolini}, Y., {Serpico}, P.~D., {Boudaud}, M., {et~al.} 2017, \prl, 119,
  241101

\bibitem[{{Giacinti} \& {Semikoz}(2023)}]{2023arXiv230510251G}
{Giacinti}, G., \& {Semikoz}, D. 2023, arXiv e-prints, arXiv:2305.10251

\bibitem[{{Ginzburg} \& {Syrovatskii}(1964)}]{1964ocr..book.....G}
{Ginzburg}, V.~L., \& {Syrovatskii}, S.~I. 1964, {The Origin of Cosmic Rays}
  (New York: Macmillan)

\bibitem[{{Gleeson} \& {Axford}(1968)}]{1968ApJ...154.1011G}
{Gleeson}, L.~J., \& {Axford}, W.~I. 1968, \apj, 154, 1011

\bibitem[{{Gordon} \& {Burton}(1976)}]{1976ApJ...208..346G}
{Gordon}, M.~A., \& {Burton}, W.~B. 1976, \apj, 208, 346

\bibitem[{{Guo} \& {Yuan}(2018)}]{2018PhRvD..97f3008G}
{Guo}, Y.-Q., \& {Yuan}, Q. 2018, \prd, 97, 063008

\bibitem[{{H{\"o}randel}(2004)}]{2004APh....21..241H}
{H{\"o}randel}, J.~R. 2004, Astroparticle Physics, 21, 241

\bibitem[{{Hunter} {et~al.}(1997){Hunter}, {Bertsch}, {Catelli}, {Dame},
  {Digel}, {Dingus}, {Esposito}, {Fichtel}, {Hartman}, {Kanbach}, {Kniffen},
  {Lin}, {Mayer-Hasselwander}, {Michelson}, {von Montigny}, {Mukherjee},
  {Nolan}, {Schneid}, {Sreekumar}, {Thaddeus}, \&
  {Thompson}}]{1997ApJ...481..205H}
{Hunter}, S.~D., {Bertsch}, D.~L., {Catelli}, J.~R., {et~al.} 1997, \apj, 481,
  205

\bibitem[{{IceCube Collaboration} {et~al.}(2023){IceCube Collaboration},
  {Abbasi}, {Ackermann}, {Adams}, {Aguilar}, {Ahlers}, {Ahrens}, {Alameddine},
  {Alves}, {et~al.}}]{IceCube2023}
{IceCube Collaboration}, {Abbasi}, R., {Ackermann}, M., {et~al.} 2023, Science,
  380, 1338

\bibitem[{{Kachelrie{\ss}} {et~al.}(2019){Kachelrie{\ss}}, {Moskalenko}, \&
  {Ostapchenko}}]{2019CoPhC.24506846K}
{Kachelrie{\ss}}, M., {Moskalenko}, I.~V., \& {Ostapchenko}, S. 2019, Computer
  Physics Communications, 245, 106846

\bibitem[{{Kafexhiu} {et~al.}(2014){Kafexhiu}, {Aharonian}, {Taylor}, \&
  {Vila}}]{2014PhRvD..90l3014K}
{Kafexhiu}, E., {Aharonian}, F., {Taylor}, A.~M., \& {Vila}, G.~S. 2014, \prd,
  90, 123014

\bibitem[{{Kamae} {et~al.}(2006){Kamae}, {Karlsson}, {Mizuno}, {Abe}, \&
  {Koi}}]{2006ApJ...647..692K}
{Kamae}, T., {Karlsson}, N., {Mizuno}, T., {Abe}, T., \& {Koi}, T. 2006, \apj,
  647, 692

\bibitem[{{Koldobskiy} {et~al.}(2021{\natexlab{a}}){Koldobskiy},
  {Kachelrie{\ss}}, {Lskavyan}, {Neronov}, {Ostapchenko}, \&
  {Semikoz}}]{2021PhRvD.104l3027K}
{Koldobskiy}, S., {Kachelrie{\ss}}, M., {Lskavyan}, A., {et~al.}
  2021{\natexlab{a}}, \prd, 104, 123027

\bibitem[{{Koldobskiy} {et~al.}(2021{\natexlab{b}}){Koldobskiy}, {Neronov}, \&
  {Semikoz}}]{2021PhRvD.104d3010K}
{Koldobskiy}, S., {Neronov}, A., \& {Semikoz}, D. 2021{\natexlab{b}}, \prd,
  104, 043010

\bibitem[{{Linden} \& {Buckman}(2018)}]{2018PhRvL.120l1101L}
{Linden}, T., \& {Buckman}, B.~J. 2018, \prl, 120, 121101

\bibitem[{{Lipari} \& {Vernetto}(2018)}]{2018PhRvD..98d3003L}
{Lipari}, P., \& {Vernetto}, S. 2018, \prd, 98, 043003

\bibitem[{{Liu} \& {Wang}(2021)}]{2021ApJ...914L...7L}
{Liu}, R.-Y., \& {Wang}, X.-Y. 2021, \apjl, 914, L7

\bibitem[{{Liu} {et~al.}(2019){Liu}, {Guo}, \& {Yuan}}]{2019JCAP...10..010L}
{Liu}, W., {Guo}, Y.-Q., \& {Yuan}, Q. 2019, \jcap, 10, 010

\bibitem[{{Ma} {et~al.}(2023){Ma}, {Xu}, {Yuan}, {Bi}, {Fan}, {Moskalenko}, \&
  {Yue}}]{2023FrPhy..1844301M}
{Ma}, P.-X., {Xu}, Z.-H., {Yuan}, Q., {et~al.} 2023, Frontiers of Physics, 18,
  44301

\bibitem[{{Ma} {et~al.}(2022){Ma}, {Bi}, {Cao}, {Chen}, {Chen}, {Cheng},
  {Gong}, {Gu}, {He}, {Hou}, {Huang}, {Huang}, {Liu}, {Shchegolev}, {Sheng},
  {Stenkin}, {Wu}, {Wu}, {Wu}, {Xiao}, {Yao}, {Zhang}, {Zhang}, \&
  {Zuo}}]{2022ChPhC..46c0001M}
{Ma}, X.-H., {Bi}, Y.-J., {Cao}, Z., {et~al.} 2022, Chinese Physics C, 46,
  030001

\bibitem[{{Marinos} {et~al.}(2023){Marinos}, {Rowell}, {Porter}, \&
  {J{\'o}hannesson}}]{2023MNRAS.518.5036M}
{Marinos}, P.~D., {Rowell}, G.~P., {Porter}, T.~A., \& {J{\'o}hannesson}, G.
  2023, \mnras, 518, 5036

\bibitem[{{Moskalenko} \& {Strong}(1998)}]{1998ApJ...493..694M}
{Moskalenko}, I.~V., \& {Strong}, A.~W. 1998, \apj, 493, 694

\bibitem[{{Panov} {et~al.}(2007){Panov}, {Adams}, {Ahn}, {Batkov},
  {Bashindzhagyan}, {Watts}, {Wefel}, {Wu}, {Ganel}, {Guzik}, {Gunashingha},
  {Zatsepin}, {Isbert}, {Kim}, {Christl}, {Kouznetsov}, {Panasyuk}, {Seo},
  {Sokolskaya}, {Chang}, {Schmidt}, \& {Fazely}}]{2007BRASP..71..494P}
{Panov}, A.~D., {Adams}, Jr., J.~H., {Ahn}, H.~S., {et~al.} 2007, Bull. Russ.
  Acad. Sci. Phys., 71, 494

\bibitem[{{Porter} \& {Strong}(2005)}]{2005ICRC....4...77P}
{Porter}, T.~A., \& {Strong}, A.~W. 2005, in International Cosmic Ray
  Conference, Vol.~4, 29th International Cosmic Ray Conference (ICRC29), Volume
  4, 77

\bibitem[{{Prodanovi{\'c}} {et~al.}(2007){Prodanovi{\'c}}, {Fields}, \&
  {Beacom}}]{2007APh....27...10P}
{Prodanovi{\'c}}, T., {Fields}, B.~D., \& {Beacom}, J.~F. 2007, Astroparticle
  Physics, 27, 10

\bibitem[{{Qiao} {et~al.}(2022){Qiao}, {Liu}, {Zhao}, {Bi}, \&
  {Guo}}]{2022FrPhy..1744501Q}
{Qiao}, B.-Q., {Liu}, W., {Zhao}, M.-J., {Bi}, X.-J., \& {Guo}, Y.-Q. 2022,
  Frontiers of Physics, 17, 44501

\bibitem[{{Reichherzer} {et~al.}(2022){Reichherzer}, {Merten}, {D{\"o}rner},
  {Becker Tjus}, {Pueschel}, \& {Zweibel}}]{2022SNAS....4...15R}
{Reichherzer}, P., {Merten}, L., {D{\"o}rner}, J., {et~al.} 2022, SN Applied
  Sciences, 4, 15

\bibitem[{{Remy} {et~al.}(2017){Remy}, {Grenier}, {Marshall}, \&
  {Casandjian}}]{2017A&A...601A..78R}
{Remy}, Q., {Grenier}, I.~A., {Marshall}, D.~J., \& {Casandjian}, J.~M. 2017,
  \aap, 601, A78

\bibitem[{{Schwefer} {et~al.}(2022){Schwefer}, {Mertsch}, \&
  {Wiebusch}}]{2022arXiv221115607S}
{Schwefer}, G., {Mertsch}, P., \& {Wiebusch}, C. 2022, arXiv e-prints,
  arXiv:2211.15607

\bibitem[{{Seo} \& {Ptuskin}(1994)}]{1994ApJ...431..705S}
{Seo}, E.~S., \& {Ptuskin}, V.~S. 1994, \apj, 431, 705

\bibitem[{{Strong} \& {Moskalenko}(1998)}]{1998ApJ...509..212S}
{Strong}, A.~W., \& {Moskalenko}, I.~V. 1998, \apj, 509, 212

\bibitem[{{Strong} {et~al.}(2007){Strong}, {Moskalenko}, \&
  {Ptuskin}}]{2007ARNPS..57..285S}
{Strong}, A.~W., {Moskalenko}, I.~V., \& {Ptuskin}, V.~S. 2007, Annu. Rev.
  Nucl. Part. Sci., 57, 285

\bibitem[{{Strong} {et~al.}(2000){Strong}, {Moskalenko}, \&
  {Reimer}}]{2000ApJ...537..763S}
{Strong}, A.~W., {Moskalenko}, I.~V., \& {Reimer}, O. 2000, \apj, 537, 763

\bibitem[{{Strong} {et~al.}(2004){Strong}, {Moskalenko}, \&
  {Reimer}}]{2004ApJ...613..962S}
---. 2004, \apj, 613, 962

\bibitem[{{Sveshnikova} {et~al.}(2013){Sveshnikova}, {Strelnikova}, \&
  {Ptuskin}}]{2013APh....50...33S}
{Sveshnikova}, L.~G., {Strelnikova}, O.~N., \& {Ptuskin}, V.~S. 2013,
  Astroparticle Physics, 50, 33

\bibitem[{{Tibaldo} {et~al.}(2021){Tibaldo}, {Gaggero}, \&
  {Martin}}]{2021Univ....7..141T}
{Tibaldo}, L., {Gaggero}, D., \& {Martin}, P. 2021, Universe, 7, 141

\bibitem[{{Tomassetti}(2012)}]{2012ApJ...752L..13T}
{Tomassetti}, N. 2012, \apjl, 752, L13

\bibitem[{{Trotta} {et~al.}(2011){Trotta}, {J{\'o}hannesson}, {Moskalenko},
  {Porter}, {Ruiz de Austri}, \& {Strong}}]{2011ApJ...729..106T}
{Trotta}, R., {J{\'o}hannesson}, G., {Moskalenko}, I.~V., {et~al.} 2011, \apj,
  729, 106

\bibitem[{{Vecchiotti} {et~al.}(2022){Vecchiotti}, {Zuccarini}, {Villante}, \&
  {Pagliaroli}}]{2022ApJ...928...19V}
{Vecchiotti}, V., {Zuccarini}, F., {Villante}, F.~L., \& {Pagliaroli}, G. 2022,
  \apj, 928, 19

\bibitem[{{Vladimirov} {et~al.}(2012){Vladimirov}, {J{\'o}hannesson},
  {Moskalenko}, \& {Porter}}]{2012ApJ...752...68V}
{Vladimirov}, A.~E., {J{\'o}hannesson}, G., {Moskalenko}, I.~V., \& {Porter},
  T.~A. 2012, \apj, 752, 68

\bibitem[{{Yan} \& {Liu}(2023)}]{2023arXiv230412574Y}
{Yan}, K., \& {Liu}, R.-Y. 2023, arXiv e-prints, arXiv:2304.12574

\bibitem[{{Yan} {et~al.}(2023){Yan}, {Liu}, {Zhang}, {Li}, {Yuan}, \&
  {Wang}}]{2023arXiv230712363Y}
{Yan}, K., {Liu}, R.-Y., {Zhang}, R., {et~al.} 2023, arXiv e-prints,
  arXiv:2307.12363

\bibitem[{{Yang} \& {Aharonian}(2019)}]{2019PhRvD.100f3020Y}
{Yang}, R., \& {Aharonian}, F. 2019, \prd, 100, 063020

\bibitem[{{Yang} {et~al.}(2016){Yang}, {Aharonian}, \&
  {Evoli}}]{2016PhRvD..93l3007Y}
{Yang}, R., {Aharonian}, F., \& {Evoli}, C. 2016, \prd, 93, 123007

\bibitem[{{Yoon} {et~al.}(2017){Yoon}, {Anderson}, {Barrau}, {Conklin},
  {Coutu}, {Derome}, {Han}, {Jeon}, {Kim}, {Kim}, {Lee}, {Lee}, {Lee}, {Lee},
  {Link}, {Menchaca-Rocha}, {Mitchell}, {Mognet}, {Nutter}, {Park},
  {Picot-Clemente}, {Putze}, {Seo}, {Smith}, \& {Wu}}]{2017ApJ...839....5Y}
{Yoon}, Y.~S., {Anderson}, T., {Barrau}, A., {et~al.} 2017, \apj, 839, 5

\bibitem[{{Yuan}(2019)}]{2019SCPMA..6249511Y}
{Yuan}, Q. 2019, Science China Physics, Mechanics, and Astronomy, 62, 49511

\bibitem[{{Yuan} \& {Feng}(2018)}]{2018SCPMA..61j1002Y}
{Yuan}, Q., \& {Feng}, L. 2018, Science China Physics, Mechanics, and
  Astronomy, 61, 101002

\bibitem[{{Yuan} {et~al.}(2020){Yuan}, {Zhu}, {Bi}, \&
  {Wei}}]{2020JCAP...11..027Y}
{Yuan}, Q., {Zhu}, C.-R., {Bi}, X.-J., \& {Wei}, D.-M. 2020, \jcap, 2020, 027

\bibitem[{{Zhang} {et~al.}(2006){Zhang}, {Bi}, \& {Hu}}]{2006A&A...449..641Z}
{Zhang}, J.-L., {Bi}, X.-J., \& {Hu}, H.-B. 2006, \aap, 449, 641

\bibitem[{{Zhang} {et~al.}(2021){Zhang}, {Qiao}, {Liu}, {Cui}, {Yuan}, \&
  {Guo}}]{2021JCAP...05..012Z}
{Zhang}, P.-P., {Qiao}, B.-Q., {Liu}, W., {et~al.} 2021, \jcap, 2021, 012

\bibitem[{{Zhang} {et~al.}(2022){Zhang}, {Qiao}, {Yuan}, {Cui}, \&
  {Guo}}]{2022PhRvD.105b3002Z}
{Zhang}, P.-P., {Qiao}, B.-Q., {Yuan}, Q., {Cui}, S.-W., \& {Guo}, Y.-Q. 2022,
  \prd, 105, 023002

\bibitem[{{Zhao} {et~al.}(2021){Zhao}, {Fang}, \& {Bi}}]{2021PhRvD.104l3001Z}
{Zhao}, M.-J., {Fang}, K., \& {Bi}, X.-J. 2021, \prd, 104, 123001

\end{thebibliography}

\clearpage
\appendix

\setcounter{figure}{0}
\renewcommand\thefigure{A\arabic{figure}}
\setcounter{table}{0}
\renewcommand\thetable{A\arabic{table}}

\section{Fermi-LAT diffuse emission without masks}

\begin{figure}[!htb]
\centering
\includegraphics[width=0.48\textwidth]{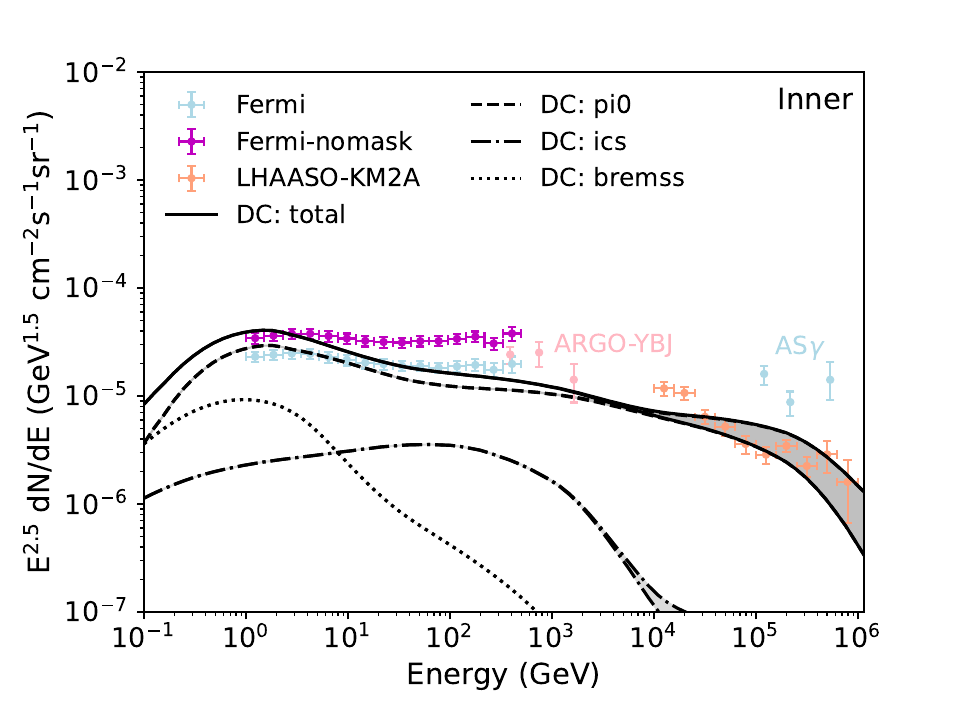}
\includegraphics[width=0.48\textwidth]{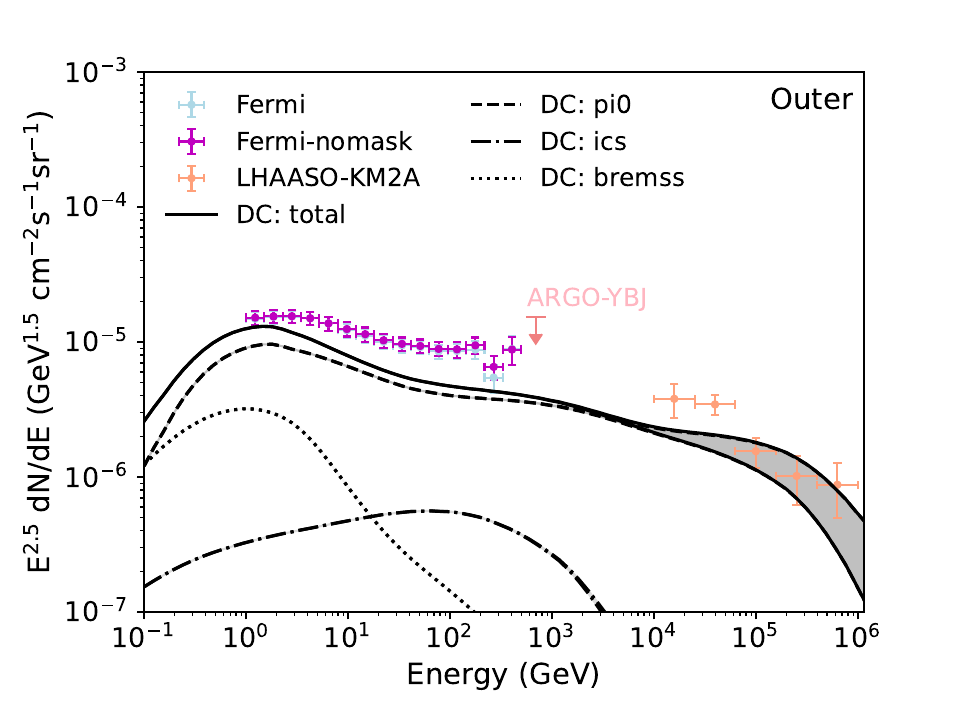}
\caption{Wide-band spectra of the diffuse $\gamma$-ray emission for the DC propagation 
model without masks, compared with Fermi-LAT data (magenta points). The labels of different
lines are the same as those in Fig.~\ref{fig:diffuse}. The other data with masks are also
shown for comparison.}
\label{fig:nomask_model}
\end{figure}

The fluxes of diffuse $\gamma$ rays in the inner and outer regions without masking
sky regions around VHE sources are given in Table \ref{tab:Fermi_flux2}. The comparison
with the DC model predictions are shown in Fig.~\ref{fig:nomask_model}.

\begin{table}[!htb]
\centering
\caption{Fluxes with 1$\sigma$ uncertainties of the Galactic diffuse emission in the 
inner and outer Galaxy regions measured by Fermi-LAT without masks.}
\begin{tabular}{cccc}
\hline\hline
$\log(E/$GeV) &  $E$ & $\phi_{\rm inner}\pm\sigma_{\rm inner}$  &    $\phi_{\rm outer}\pm\sigma_{\rm outer}$        \\
                        & (GeV)              & (GeV$^{-1}$cm$^{-2}$s$^{-1}$sr$^{-1}$) & (GeV$^{-1}$cm$^{-2}$s$^{-1}$sr$^{-1}$) \\ \hline
0.00-0.18                 & 1.23              & $(2.05 \pm 0.22 ) \times 10^{-5}$    &   $(9.03 \pm 1.04 ) \times 10^{-6}$         \\
0.18-0.36                 & 1.86               & $(7.67 \pm 0.80 ) \times 10^{-6}$    &    $(3.27 \pm 0.35 ) \times 10^{-6}$        \\
0.36-0.54                 & 2.82               & $(2.84 \pm 0.30 ) \times 10^{-6}$     &   $(1.17 \pm 0.13 ) \times 10^{-6}$        \\
0.54-0.72                 & 4.26               & $(1.01 \pm 0.11 ) \times 10^{-6}$     &   $(3.99 \pm 0.45 ) \times 10^{-7}$        \\
0.72-0.90                 & 6.45               & $(3.40 \pm 0.38 ) \times 10^{-7}$      &  $(1.30 \pm 0.15 ) \times 10^{-7}$        \\
0.90-1.08                 & 9.76             & $(1.14 \pm 0.13) \times 10^{-7}$       &   $(4.19 \pm 0.50 ) \times 10^{-8}$       \\
1.08-1.26                 & 14.78             & $(3.86 \pm 0.44) \times 10^{-8}$       &    $(1.36 \pm 0.16 ) \times 10^{-8}$   \\
1.26-1.44                 & 22.36              & $(1.34  \pm 0.15) \times 10^{-8}$       &   $(4.35 \pm 0.53 ) \times 10^{-9}$   \\
1.44-1.62                 & 33.84              & $(4.70 \pm 0.51) \times 10^{-9}$       &    $(1.45 \pm 0.17 ) \times 10^{-9}$   \\
1.62-1.80                 & 51.21              & $(1.71 \pm 0.18) \times10^{-9}$        &    $(4.98 \pm 0.58 ) \times 10^{-10}$   \\ 
1.80-1.98                 & 77.50              & $(6.10\pm 0.65) \times10^{-10}$        &    $(1.68 \pm 0.20 ) \times 10^{-10}$    \\ 
1.98-2.16                 & 117.28              & $(2.27 \pm 0.25) \times10^{-10}$        &    $(5.92 \pm 0.77 ) \times 10^{-11}$    \\ 
2.16-2.34                 & 177.48              & $(8.46  \pm 0.96) \times10^{-11}$        &    $(2.25 \pm 0.32 ) \times 10^{-11}$   \\ 
2.34-2.52                 & 268.58              & $(2.59  \pm 0.33) \times10^{-11}$        &   $(5.53 \pm 1.14 ) \times 10^{-12}$     \\ 
2.52-2.70                 & 406.45             & $(1.14  \pm 0.16) \times10^{-11}$       &    $(2.64 \pm 0.61 ) \times 10^{-12}$ \\ \hline
\end{tabular}
\label{tab:Fermi_flux2}
\end{table}

\section{Fitting results of primary and secondary spectra of CRs}

\begin{figure}[!htb]
\centering
\includegraphics[width=0.7\columnwidth]{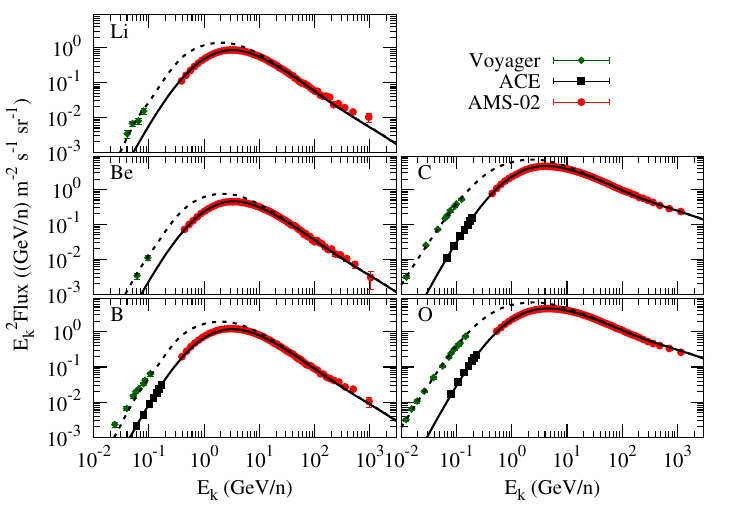}
\includegraphics[width=0.7\columnwidth]{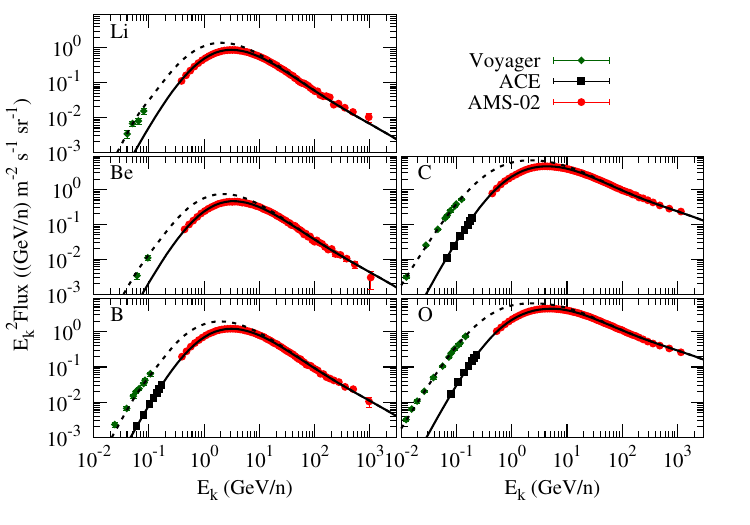}
\caption{Best-fitting results for spectra of Li, Be, B, C, and O nuclei,
for the DC model (top panels) and DR model (bottom panels), compared with 
the measurements \citep{2017PhRvL.119y1101A,2018PhRvL.120b1101A,
2016ApJ...831...18C,2019SCPMA..6249511Y}. 
}
\label{fig:nuclei}
\end{figure}

\end{document}